\documentclass[table,twocolumn]{aastex62}
\usepackage{amstext}
\usepackage{amsmath}
\usepackage{apjfonts}
\usepackage[capbesideposition={top}]{floatrow}
\usepackage{float}
\usepackage{cancel}
\usepackage{floatrow}

\newcommand{\beqar}{\begin{eqnarray}}
\newcommand{\eeqar}{\end{eqnarray}}

\newcommand{\beq}{\begin{equation}}
\newcommand{\eeq}{\end{equation}}

\definecolor{nick}{HTML}{006400}

\begin{document}
\title{Extreme scale height variations and nozzle shocks in warped disks}


\correspondingauthor{Nick Kaaz}
\email{nkaaz@u.northwestern.edu}

\author[0000-0002-5375-8232]{Nicholas Kaaz}
\affiliation{Department of Physics \& Astronomy, Northwestern University, Evanston, IL 60202, USA}
\affiliation{Center for Interdisciplinary Exploration \& Research in Astrophysics (CIERA), Evanston, IL 60202, USA}

\author{Yoram Lithwick}
\affiliation{Department of Physics \& Astronomy, Northwestern University, Evanston, IL 60202, USA}
\affiliation{Center for Interdisciplinary Exploration \& Research in Astrophysics (CIERA), Evanston, IL 60202, USA}

\author[0000-0003-4475-9345]{Matthew Liska}
\affiliation{Center for Relativistic Astrophysics, Georgia Institute of Technology, Howey Physics Bldg, 837 State St NW, Atlanta, GA 30332, USA}

\author[0000-0002-9182-2047]{Alexander Tchekhovskoy}
\affiliation{Department of Physics \& Astronomy, Northwestern University, Evanston, IL 60202, USA}
\affiliation{Center for Interdisciplinary Exploration \& Research in Astrophysics (CIERA), Evanston, IL 60202, USA}





\begin{abstract} 
Accretion disks around both stellar-mass and supermassive black holes are likely often
warped. Whenever a disk is warped, its scale height varies with azimuth. Sufficiently strong warps cause extreme compressions of the scale height, which fluid parcels ``bounce'' off of twice per orbit to high latitudes. We study the dynamics of strong warps using: (i) the nearly-analytic ``ring theory'' of \cite{fairbairn_ogilvie_2021a}, which we generalize to the Kerr metric; and (ii) 3D general-relativistic hydrodynamic simulations of tori (``rings'') around black holes, using the \verb|H-AMR| code. We initialize a ring with a warp and study its evolution on tens of orbital periods. The simulations agree excellently with the ring theory until the warp amplitude, $\psi$, reaches a critical value $\psi_{\rm c}$. When $\psi>\psi_{\rm c}$, the rings enter the bouncing regime. We analytically derive (and numerically validate) that $\psi_{\rm c}\approx (r/r_{\rm g})^{-1/2}$ in the non-Keplerian regime, where $r_{\rm g}=GM/c^2$ is the gravitational radius and $M$ is the mass of the central object. Whenever the scale height bounces, the vertical velocity becomes supersonic, leading to ``nozzle shocks'' as gas collides at the scale height minima.
Nozzle shocks damp the warp within $\approx10-20$ orbits, which is not captured by the ring theory. Nozzle shock dissipation leads to inflow timescales 1-2 orders of magnitude shorter than unwarped $\alpha$ disks which may result in rapid variability, such as in changing-look active galactic nuclei or in the soft state of X-ray binaries. We propose that steady disks with strong warps may self-regulate to have amplitudes near $\psi_{\rm c}$. 

\end{abstract}


\section{Introduction}
\label{sec:intro}

When gas falls within the sphere of influence of a gravitating object -- be it a star, a compact object, or maybe a binary system -- the angular momentum of the gas is not necessarily aligned with the angular momentum of the gravitating object(s). These misaligned configurations tend to torque the gas such that the resulting accretion disk has a radially-dependent orientation.  Such disks are called ``warped'' and can have dramatically different behavior than flat, planar disks. In this work, we are interested in the evolution of warps around black holes (BHs), although our results also carry over to other systems such as protoplanetary disks. 

There is observational evidence for warps in accretion disks around both supermassive and stellar mass BHs. In the former case, there are observations of sub-parsec scale maser emission within active galactic nuclei (AGN) that are consistent with warps \citep{greenhill_2003,zhao_2018,zaw_2020}, most notably in NGC 4258 \citep[]{miyoshi_1995,greenhill_1995}. The latter case is evidenced by a large population of X-ray binaries (XRBs) which exhibit months-to-years long periodicity. This is much longer than the orbital timescale and has often been argued to be a signature of warp-induced precession \citep{priedhorsky_1987,smale_1992,ogilvie_dubus_2001,kotze_2012}. XRBs also exhibit quasi-periodic oscillations which may be driven by precession as well \citep[][]{stella_vietri_1998,ingram_2016,gibwa_2023,deepika_2023,deepika_2024}. In addition, some protoplanetary disks exhibit shadows that may be produced by a warp in the disk \citep{marino_perez_2015,stolker_2016,debes_2017}.  Such warps could be driven by
inclined planets 
\citep[e.g., ][]{nealon_2018,zhu_2019}.

Given that we know and expect warped disks to occur in nature, one of the jobs of theory is to be able to predict the long-term evolution of the warp. One of the earliest efforts to do so is \cite{bardeen_petterson_1975}, who pointed out that thin, viscous and slightly tilted disks accreting onto rotating BHs should gradually align with the BH spin axis up to some radius. However, the behavior of a warp depends on the disk's internal hydrodynamics \citep{papaloizou_1983}, which were not correctly modeled in the early studies. One of the main complications is that warps feature a radial variation of the vertical position of the disk midplane, allowing radial pressure gradients to drive strong, oscillatory, horizontal motions parallel to the disk surface. In Keplerian potentials, the orbital frequency equals the radial and vertical epicyclic frequencies, which causes these horizontal motions to be resonantly driven to large amplitudes. This leads to rapid communication of the warp which, in linear theory, is mediated by bending waves that propagate radially at a fraction of the sound speed \citep{papaloizou_lin_1995}. In
sufficiently non-Keplerian potentials, the degeneracy between the orbital and epicyclic frequencies is broken, and these motions are not as strongly driven. This is also true in sufficiently viscous accretion disks. This has led to a dichotomy in the analytic treatment of warped accretion disks between a ``resonant'' regime, where the potential is sufficiently close to Keplerian and the viscosity is not too strong, and the ``non-resonant'' regime \citep{papaloizou_1983}. Of the two regimes, the nonlinear, non-resonant theory is more fleshed out \citep[]{pringle_1992,ogilvie_1999,ogilvie_2000}. While the nonlinear resonant regime has received some attention \citep[e.g.,][]{ogilvie_2006,dullemond_2022}, it remains less well-understood. The recent work of \cite{fairbairn_ogilvie_2021a} took a novel approach: instead of modeling disks, they modeled radially-narrow ``rings'', which facilitated the study of resonantly driven, strongly nonlinear warps. In an accompanying work, \cite{fairbairn_ogilvie_2021b}, the authors found that above a certain warp amplitude, a nonlinear resonance condition -- separate from the aforementioned Keplerian resonance -- drives extreme scale height oscillations twice an orbit. This, in turn, is expected to lead to important dynamical consequences for the warp and may have distinct observational signatures. One of the main motivations of our work is to improve the understanding of this phenomenon. 

In parallel with these analytic calculations, there has been a growing effort to numerically model warped accretion disks. Especially in earlier works, a common tool has been smoothed-particle hydrodynamic (SPH) simulations \citep{larwood_1996,nelson_1999,lodato_2006}. SPH simulations have found good agreement with \cite{ogilvie_1999} for viscous disks \citep{lodato_2010} and have unveiled exciting new dynamics in extreme warps, such as the tearing of accretion disks into discrete planes \citep{nixon_2012,nixon_2013}. However, it is unclear how well SPH captures the internal motions of warped disks, especially when the viscosity is small. This is highlighted by \cite{deng_ogilvie_2022}, who found that Godunov-type shock capturing methods matched analytic predictions with much higher accuracy. In the last several years, magnetohydrodynamic \citep["MHD",][]{sorathia_2013,krolik_hawley_2015,hawley_krolik_2018,hawley_krolik_2019}, general-relativistic MHD \citep["GRMHD",][]{liska_2018,liska_2019,liska_2021,white_2019, white_2020, kaaz_2023}, and radiation GRMHD \citep{liska_kaaz_2023} simulations have also become sufficiently sophisticated to simulate warped accretion disks around rapidly rotating BHs. These simulations have confirmed various dynamical phenomena, including jet precession \citep{liska_2018}, Bardeen-Petterson alignment in mildly tilted disks \citep{liska_2019}, and the tearing of more strongly tilted disks \citep{liska_2021,HAMR,liska_kaaz_2023}. Closely related to this work is \cite{kaaz_2023} who found that warped GRMHD disks undergo extreme scale height oscillations twice an orbit, leading to ``nozzle'' shocks that drive rapid accretion and provide feedback onto the warp. Shocks were also found in \cite{held_ogilvie_2024}, wherein the authors studied strong scale height oscillations of disks more generally. This dissipation mechanism is purely hydrodynamic, suggesting that magnetic fields may not be necessary to drive accretion in strongly warped disks.

Global simulations of thin accretion disks are extremely expensive. This is due to a few reasons. Firstly, and most importantly, it is necessary to resolve small-scale turbulent eddies within the scale height, which becomes increasingly difficult the thinner the disk is. For instance, without adaptive mesh refinement, if a disk is half as thin it requires eight times as many cells and twice as many timesteps to resolve. Even with adaptive mesh refinement, reducing the disk thickness by a factor of two can still require a factor of four more resolution elements \citep[see., e.g., ][]{HAMR}. Secondly, velocities are of order the speed of light near the event horizon, so to obey the Courant condition the time-step must be small. This is particularly constraining on otherwise advantangeous spherical-polar grids due to azimuthal cell squeezing at the polar boundary \citep{ressler_2017,HAMR}. Thirdly, to achieve a steady state, it is usually desirable to evolve the simulation on the accretion timescale, which is often very long. While a handful of such simulations have been performed, it remains difficult or impossible to explore a large parameter space at sufficiently high resolution due to the computational expense involved -- the most advanced simulations currently cost tens of millions of core hours\footnote{Assuming 1 GPU-hour is $\approx20$ core hours} in compute time \citep{HAMR}. Additionally, the complicated behavior of first-principles simulations can make it difficult to extract insight. 

To make progress, here  we
intend to help bridge the gap between analytic models and global multi-physics simulations. We do this by modelling three-dimensional, hydrodynamic ring-like tori on tens of orbital timescales, which we compare with the ring theory \citep{fairbairn_ogilvie_2021a} in order to garner insights into more realistic warped accretion disks. Specifically, we 
explain the onset of resonant bouncing and nozzle shocks in strongly warped, general-relativistic disks. In Section \ref{sec:ring}, we review the ring theory for warped disks, describe its extension to the Kerr metric, and describe the dynamics of small warps. In Section \ref{sec:approach}, we introduce and validate our numerical approach. In Section \ref{sec:bouncing_regime}, we describe our results on the onset of bouncing and the accompanying nozzle shocks. In Section \ref{sec:discussion}, we discuss the application to fully global, turbulent accretion disks and summarize our main findings.  

\begin{figure*}
    \centering
    \includegraphics[width=\textwidth]{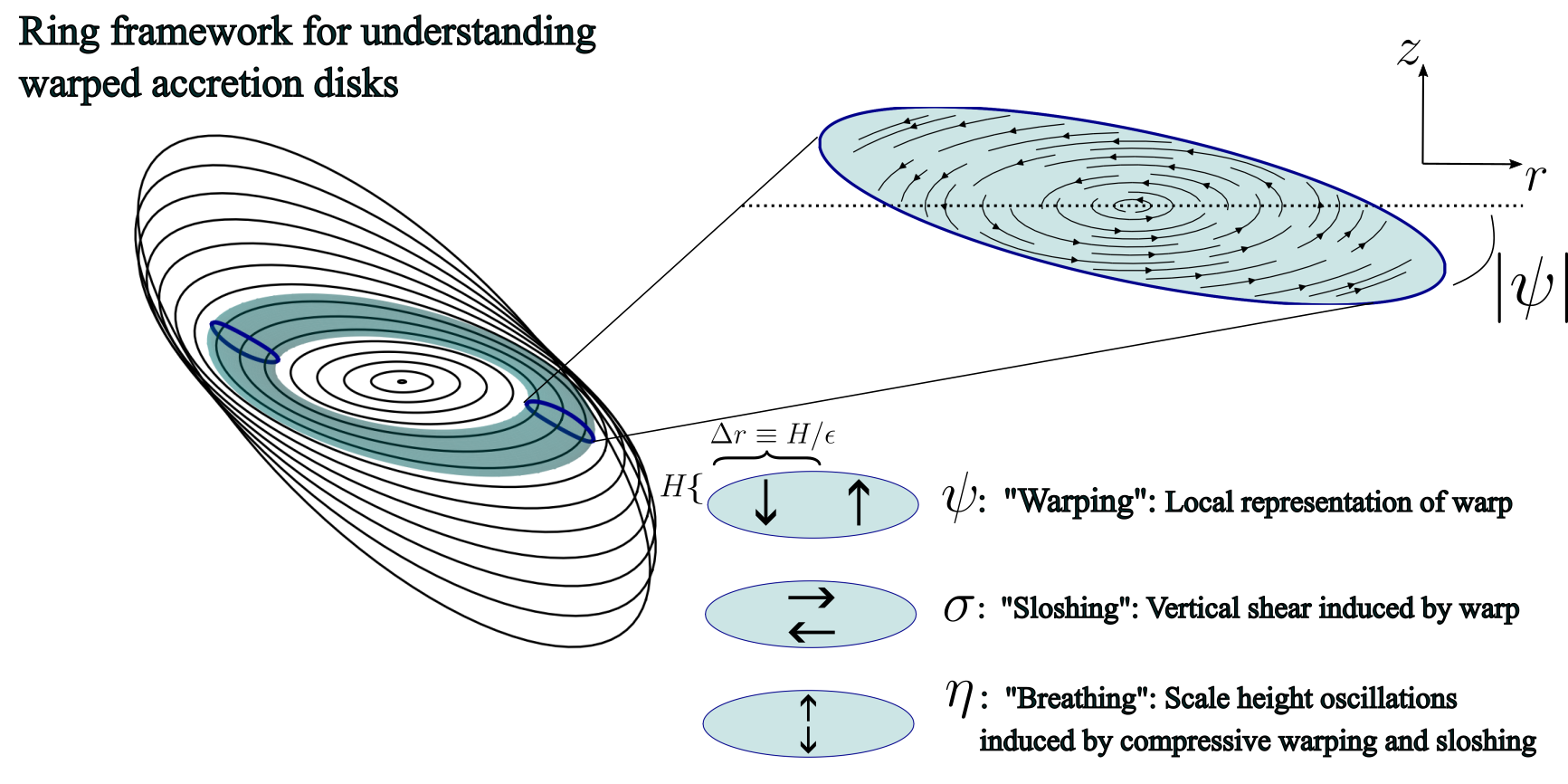}
    \caption{We can model a radially-narrow section of a warped accretion disk as a ``ring'', shown in teal in this cartoon. \textbf{Left.} We depict a warped disk with a series of concentric annuli (black) with radially-dependent orientations. The teal torus shows a radially-narrow section of the global disk, which we call a ``ring''. We also draw dark-blue contours around this torus to highlight examples of its cross-section half an orbital period apart. The ring theory describes how this cross-section evolves in the reference frame that co-rotates with the orbit. \textbf{Top right.} Zoom-in on the ring cross-section, which is assumed to be elliptical. In the rotating reference frame, a warped ring is represented by the periodic tilting of its elliptical contours. The magnitude of the tilting angle is equal to the warp amplitude, $|\psi|$. Within the cross-section, we have drawn velocity streamlines corresponding to a linear tilting mode (Eq.~\ref{eq:app:warping_eigenvectors}). \textbf{Bottom right.} The linear tilting mode is described by both a ``warping'' motion ($\psi$) associated with the tilting of elliptical contours and a ``sloshing'' motion ($\sigma$) describing the vertical shear of the radial velocity. At higher order, the warping and sloshing motions multiply  to drive the vertical ``breathing'' ($\eta$) of the ring. Here we have also labeled the half-thickness $H$ and the half-width $\Delta r\equiv H/\epsilon$ of the ring, where $\epsilon$ is the height-to-width aspect ratio.}
\label{fig:ring_diagram}
\end{figure*}

\section{Ring theory}
\label{sec:ring}
In this work, we study general-relativistic warps in thin, radially-narrow tori (``rings''), as depicted in Figure \ref{fig:ring_diagram}. We want to connect the rich phenomenology found in 3D simulations of global, general relativistic warped accretion \citep[][``K23'']{kaaz_2023} with the body of analytic work which, while powerful in its ability to model nonlinear warps, is limited in the complexity of the systems it can probe. We follow the framework devised by \cite{fairbairn_ogilvie_2021a} (``FO21a''), who derived equations for oscillating rings and found close correspondence with the theory of linear and nonlinear bending waves \citep[][]{ogilvie_2006,ogilvie_latter_2013}. We describe the ring theory and its application to the Kerr metric in detail in Appendix \ref{app:ring_theory}. In the following sections, we provide a more qualitative overview.  

\subsection{Summary of the Fairbairn \& Ogilvie ring model}

 Here, we summarize FO21a's main assumptions:
\begin{itemize}
    \item The ring theory is derived in the local shearing sheet approximation \citep{hawley_1995}. That is, FO21a expand the equations of ideal, compressible hydrodynamics in the neighborhood of a circular orbit located at radius $r_0$ and orbiting about a central mass $M$ at the local Keplerian frequency $\Omega$. The gas obeys a polytropic equation of state $p=K\rho^\gamma$, where $p$ is the gas pressure, $\rho$ is the gas density, $K$ is a constant and $\gamma$ is the adiabatic index. 
    
    \item Within the local shearing sheet, FO21a derive hydrodynamic, ring-like equilibria. These rings are described by an aspect ratio, 
    \begin{equation}
        \epsilon = H/\Delta r,
        \label{eq:aspect_ratio}
    \end{equation}
    where $H$ is the scale height at the center of the ring and $\Delta r$ is the half-width. The rings are in vertical hydrostatic equilibrium. They are radially confined by assuming an orbital shear that is slightly greater than the shear of circular orbits, which naturally leads to a tapering of the density contours in the radial directions.

    \item The rings then need to be able to warp. For this, FO21a generalize their equilibrium solution to one with a linear flow field,
    \begin{equation}
        u_i = A_{ij}x_j,
        \label{eq:flow_matrix}
    \end{equation}
    where $u_i$ is the gas velocity and $A_{ij}$ is a time-dependent, $y$-independent ($A_{i2}=0$) ``flow matrix'' and $x_i=(x,y,z)$ where Latin indices obey the convention $i=1,2,3$.  The local Cartesian coordinates correspond to spherical coordinates via the relations $x=r-r_0$, $y=r_0(\varphi-\Omega t)$ and $z=r_0{\rm sin}\theta$. The orbital shear is described by $A_{21}$, which is the only non-zero component of $A_{ij}$ in the equilibrium state. Other components of $A_{ij}$ describe how the rings warp and jiggle. 
    The rings are always composed of elliptical density contours in the $r$-$z$ plane, which this flow field distorts. We will describe how $A_{ij}$ connects to warps shortly.

\end{itemize}

From these assumptions, FO21a derived a set of ten coupled, nonlinear ordinary differential equations describing the time-evolution of oscillating rings. These equations describe the evolution of (i) the six non-zero components of the flow matrix $A_{ij}$, (ii) the three quantities defining the shape of the elliptical density contours, parameterized by a matrix $S_{ij}$ (Eq.~\ref{eq:app:shape_matrix}), and (iii) the temperature of the gas. We derive these ordinary differential equations in Appendix \ref{app:ring_theory}, generalizing  F021a's derivation to the Kerr metric. The result is Equations \ref{eq:fa:1}-\ref{eq:fa:10}.

\subsection{Application to the Kerr metric}
    
The Kerr metric ring equations are identical to the Newtonian case except that the orbital shear, $S$ (Eq.~\ref{eq:app:orbital_shear}), and the fluid-frame vertical epicyclic frequency, $\nu$ (Eq.~\ref{eq:app:ff_vertical_epicyclic}), are modified by relativistic correction factors. The fluid-frame radial epicyclic frequency, $\kappa$ (Eq.~\ref{eq:app:ff_radial_epicyclic}), is also important, but does not appear explicitly in the 
equations. For a BH with zero spin ($a=0$) these quantities are,
\begin{equation}
\begin{aligned}
&S\approx\frac{3}{2}(1+r_{\rm g}/2r_0)\Omega\\
&\nu\approx(1+3r_{\rm g}/2r_0)\Omega\\
&\kappa\approx(1-3r_{\rm g}/2r_0)\Omega,
\end{aligned}
\label{eq:apprx_relativistic_freqs}
\end{equation}
in the limit $r_0\gg r_g$, where $\Omega=\sqrt{GM/r_0^3}$ is the frequency at which the frame rotates and $r_{\rm g}=GM/c^2$ is the gravitational radius. These frequencies differ from the Keplerian case, where the equality of all three frequencies, $\Omega=\nu=\kappa$, allows the warp to resonantly drive horizontal velocities to large values \citep{papaloizou_1983}. 

\subsection{Ring oscillations}
\label{sec:rings:linear}


\subsubsection{Linear tilting modes}
\label{sec:rings:linear:warp}

We quantify the disk warp (left of Fig.~\ref{fig:ring_diagram})  with $|\psi|=|\frac{d\hat{l}}{d{\rm ln}r}|$, where $\hat{l}(r)$ is the angular momentum unit vector. In the orbiting frame, $\psi$ manifests as the tilting of density contours back and forth on an orbital timescale (top right of Fig.~\ref{fig:ring_diagram}). We can define a time-dependent $\psi$ locally\footnote{If the ring expands radially, then $\psi$ as defined in Eq.~\ref{eq:psidef} will deviate from $|\frac{d\hat{l}}{d{\rm ln}r}|$.},
\begin{equation}
\psi \equiv A_{31}/\nu,\quad``{\rm Warping}"
\label{eq:psidef}
\end{equation}
Whenever $\psi\neq0$, it causes a periodic reorientation of pressure gradients that make the gas slosh back and forth, 
\begin{equation}
    \sigma \equiv A_{13}/\nu,\quad``{\rm Sloshing}",
\end{equation}
where $\sigma$ is also time-dependent. Sloshing is 
associated with eccentric fluid orbits whose radial velocities change sign across the ring midplane. We introduce perturbations with time dependence ${\rm exp}({i\omega_\psi t})$ to $\psi$, $\sigma$ and the other quantities that break the midplane symmetry of the ring to derive linear ``tilting'' modes (Appendix \ref{app:linear_modes:warping}). In the lab frame, tilting modes correspond to bending waves with wavelength $\approx\Delta r$. The linearized equation of motion (Eq.~\ref{app:eq:linear_warping_eigenproblem}) is,
\begin{equation}
    \frac{d^2}{dt^2}
    \begin{pmatrix}
    \sigma\\\psi
    \end{pmatrix} +
    \begin{pmatrix}
    \kappa^2&\nu^2\\
    \nu^2\epsilon^2&\nu^2\\
    \end{pmatrix}
    \begin{pmatrix}
    \sigma\\\psi
    \end{pmatrix}=0
    \label{eq:linear_warping_eom},
\end{equation}
and the resulting eigenvalues are,
\begin{equation}
    \omega_{\psi\pm}^2 = \nu^2\left(1 - \frac{1}{2}\delta \pm \sqrt{\frac{1}{4}\delta^2 + \epsilon^2}\right),
    \label{eq:linear_warping_eigenvalues}
\end{equation}
where $\delta$ encompasses the ``non-Keplerianity'' of the disk,
\begin{equation}
    \delta \equiv 1 - \kappa^2/\nu^2 \approx 6r_{\rm g}/r_0
\label{eq:deviation_resonance},
\end{equation}
where the approximation is taken in the zero spin limit far from the BH. The eigenvalue in Equation \ref{eq:linear_warping_eigenvalues} depends on the ratio $\delta/\epsilon$. The tilting mode is
 considered to be
``non-Keplerian'' when $\delta\gg\epsilon$.  If we generalize to a disk by taking $\Delta r\rightarrow r$ in Eq.~\ref{eq:aspect_ratio}, then $\epsilon\sim H/r$. In thin rings/disks with $\epsilon\sim H/r\sim0.01$, the tilting mode is non-Keplerian if $r<600r_{\rm g}$. This is remarkably far from the event horizon -- most GR effects can be neglected at much smaller radii. Throughout this paper, we focus on the non-Keplerian limit. The eigenvalues in the non-Keplerian limit are,
\begin{equation}
\begin{aligned}
    \omega_{\psi,+}^2 &\approx \nu^2+\frac{2\epsilon^2\nu^2}{\delta^2},\quad``{u_z-{\rm dominated}}" \\
    \omega_{\psi,-}^2 &\approx \kappa^2 - \frac{2\epsilon^2\nu^2}{\delta^2},\quad``{u_x-{\rm dominated}}"
    \label{eq:apprx_bending_eigenvalues}
\end{aligned}
\end{equation}
The associated pair of eigenvectors is,
\begin{equation}
\begin{aligned}
    \begin{pmatrix}
    \sigma\\\psi
    \end{pmatrix}
    \approx\,
    &c_+
    \begin{pmatrix}
    1/\delta\\1
    \end{pmatrix}{\rm exp}(i\omega_{\psi,+}t)
    \\+&c_-
    \begin{pmatrix}
    1/\epsilon^2\\1
    \end{pmatrix}{\rm exp}(i\omega_{\psi,-}t),
    \label{eq:linear_warping_eigenvectors}
\end{aligned}
\end{equation}
and we provide the full expression in Eq.~\ref{eq:app:warping_eigenvectors}.
In Equation \ref{eq:apprx_bending_eigenvalues}, the ``$+$'' and ``$-$'' 
modes follow
$\nu$ and $\kappa$, and are dominated by the vertical ($u_z\propto \Delta r\psi$) and radial ($u_x\propto H\sigma$) velocities\footnote{In Keplerian disks, warping and sloshing motions are resonantly coupled, and $u_x=u_z$.}, respectively. In the ``$u_z$-dominated'' (+) mode, $u_z/u_x = \delta/\epsilon\gg 1$. In the ``$u_x$-dominated'' (-) mode, $u_z/u_x=\epsilon\ll1$. Since we care about warp-driven dynamics, we will focus on the $u_z$-dominated mode. Indeed, for the same value of $\psi$, the $u_x$-dominated mode has a factor $\approx \delta^2/\epsilon^4$ more energy than the vertical mode, so we expect that it is very difficult to excite.

\subsubsection{Linear breathing modes}
\label{sec:rings:linear:breathing}

The vertical breathing of the ring is,
\begin{equation}
\eta \equiv A_{33}/\nu,\quad``{\rm Breathing}"
\end{equation}
There is also radial breathing, $\eta_{\rm R}\equiv A_{11}/\nu$, which we will not discuss beyond this section. We introduce perturbations with time dependence ${\rm exp}({i\omega_\eta t})$ to $\eta$, $\eta_{\rm R}$ and other related quantities to derive linear breathing modes (Appendix \ref{app:linear_modes:breathing}). The linearized equation of motion (Eq.~\ref{eq:app:linear_breathing_eigenproblem}) is,
\begin{equation}
    \frac{d^2}{dt^2}
    \begin{pmatrix}
    \eta\\\eta_{\rm R}
    \end{pmatrix} +
    \begin{pmatrix}
    \nu^2(\gamma-1)&\nu^2(\gamma+1)\\
    \kappa^2+\gamma\epsilon^2\nu^2&\epsilon^2\nu^2(\gamma-1)\\
    \end{pmatrix}
    \begin{pmatrix}
    \eta\\\eta_{\rm R}
    \end{pmatrix}=0
    \label{eq:linear_breathing_eigenproblem},
\end{equation}
and the resulting eigenvalues in the thin ($\epsilon\rightarrow0$) ring limit are,
\begin{equation}
\begin{aligned}
    &\omega_{\eta,+}^2 \approx \kappa^2+\mathcal{O}(\epsilon^2),\quad\qquad\,\,\,\,``{u_x-{\rm dominated}}"\\
    &\omega_{\eta,-}^2 \approx \nu^2(1 +\gamma)+\mathcal{O}(\epsilon^2),\quad``{u_z-{\rm dominated}}"
\end{aligned}
\label{eq:linear_breathing_eigenvalues}
\end{equation}
In the $\epsilon\rightarrow0$ limit, the ring becomes a (nearly) infinite disk, and radial pressure gradients play a negligible role. This means $\eta$ and $\eta_{\rm R}$ couple weakly. The ``+'' and ``-'' are dominated by the radial ($u_x\propto \eta_{\rm R}\Delta r$) and vertical ($u_z\propto \eta H$) velocities, respectively. In the ``$u_x$-dominated'' (+) mode, $u_z/u_x$ is $\mathcal{O}(\epsilon)$. This is why the mode oscillates, to leading order, at the radial epicyclic frequency. In the ``$u_z$-dominated'' (-) mode, $u_z/u_x$ is $\mathcal{O}(\epsilon^{-1})$. This is why, to leading order, the ``$-$'' mode depends on the vertical epicyclic frequency. It also depends on $\gamma$, since both gravity and vertical pressure gradients act as restoring forces. We interpret the ``$-$'' as a scale height oscillation. Since here we focus on scale height oscillations, this is the physically relevant mode to consider.

\subsubsection{Forcing of breathing modes by tilting modes}
\label{sec:rings:linear:quasi}
Moving forward, we will assume only ``$u_z$-dominated'' tilting modes (Eq.~\ref{eq:apprx_bending_eigenvalues} and ``$u_z$-dominated'' breathing modes (Eq.~\ref{eq:linear_breathing_eigenvalues}), so we will take $\omega_{\psi,+}\rightarrow\omega_\psi$ and $\omega_{\eta,-}\rightarrow\omega_\eta$. At second order in $\psi$, tilting modes force "quasi-linear" breathing modes that approximately obey the relation (Eq.~\ref{eq:app:forced_breathing_solution}), neglecting the $u_{\rm x}$-dominated breathing mode), 
\begin{equation}
    \ddot{\eta} + (\gamma+1)\nu^2\eta =-iF_\eta{\rm exp}(i2\omega_\psi t),
\end{equation}
where (Eq.~\ref{eq:app:forcing_amplitude}),
\begin{equation}
    F_\eta \equiv \nu^2\left[(\gamma+3) - 2\delta/3(\gamma-1)\right]|\psi\sigma|,
    \label{eq:forcing_amplitude}
\end{equation}
Here, we have assumed a pure linear tilting mode and kept only leading order terms in $\epsilon$ and $\delta$. The $\propto|\psi\sigma|$ scaling can be understood
by taking the divergence of the pressureless Euler equation,
\begin{equation}
    \partial_t \vec{\nabla}\cdot\vec{u} \approx -2\psi\sigma\Omega^2,
\end{equation}
which says that fluid parcels are periodically (de)compressed by an $\mathcal{O}(\psi\sigma)$ factor. Since both $\psi$ and $\sigma$ have time dependence ${\rm exp}(i\omega_\psi t)$, compressions oscillate at frequency $2\omega_\psi\approx2\Omega$. In other words, warping and sloshing shove fluid parcels into and away from one another, forcing them to compress about twice an orbit. More generally, small breathing modes in the presence of a tilting mode can be described as a superposition of homogeneous (``natural'') and particular (``forced'') solutions,
\begin{equation}
\eta \approx \eta^{\rm (n)}{\rm exp}(i(\omega_\eta t+\phi))+\frac{F_\eta}{\omega_\psi^2-\omega_\eta^2}{\rm exp}(i2\omega_\psi t),
\label{eq:ql_superposition}
\end{equation}
where $\phi$ is the phase. This results in a forced amplitude $\eta^{\rm (f)}=F_\eta/(\omega_\psi^2-\omega_\eta^2)$ which is approximately (see Eq.~\ref{eq:app:ql_breathing}),
\begin{equation}
    \eta^{\rm (f)} \approx |\psi|^2\frac{2\gamma+6 - 4\delta/3(\gamma-1)}{\bigl(\gamma-3\bigl)\bigl(\delta+\sqrt{\delta^2 + 4\epsilon^2}\bigl)}
    \label{eq:forced_breathing}
\end{equation}
When the warp is small, forced breathing is a small correction to the flow. 
When the warp grows large, Equation \ref{eq:forced_breathing} breaks down, and $\eta$ grows large. This dramatically impacts the flow, as we will discuss in Section \ref{sec:bouncing_regime}.

\section{Approach}
\label{sec:approach}


\begin{figure*}
    \centering
    \includegraphics[width=\textwidth]{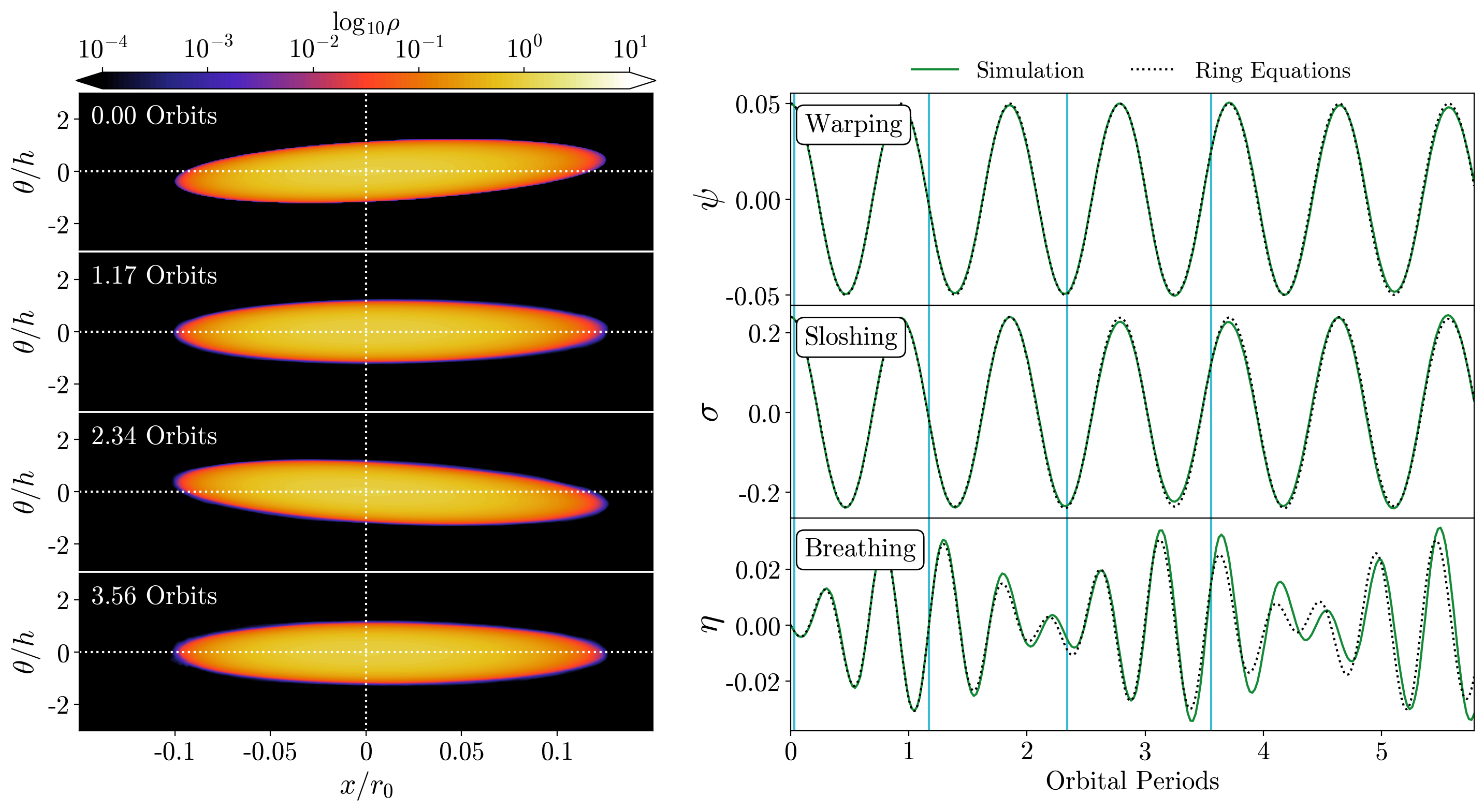}
    \caption{Comparison of simulated linear and second order quasi-linear modes to the ring theory for warped accretion disks. Narrow tori, which are not too strongly warped, gently oscillate in a manner that is reasonably approximated by the ring theory. We depict this with simulation L\_e0.1p0.05, where we have initialized a linear ``$+$'' tilting mode (Eq.~\ref{eq:linear_warping_eigenvectors}). \textbf{Left panels.}
    We depict $x-\theta$ cross-sections of the flow, where $x=r-r_0$. We set the azimuthal coordinate $\varphi=\Omega t$, such that we follow the gas in the rotating reference frame. \textbf{Right panels.} We compare the time evolution of the ``warping'' ($\psi$), ``sloshing'' ($\sigma$) and ``breathing'' ($\eta$) motions (green solid lines) of the simulation with the prediction from the ring equations (black dotted lines, Eqs. \ref{eq:fa:1}-\ref{eq:fa:10}). The breathing mode, $\eta$, is driven nonlinearly by the coupling of $\psi$ and $\sigma$ (see Section \ref{sec:rings:linear:breathing}). The beating pattern of $\eta$ results from the mixing of the forced mode and the natural mode of $\eta$ (Eq.~\ref{eq:ql_superposition}). The vertical lines indicate the times that correspond to the left panels.}
    \label{fig:gentle_ring}
\end{figure*}

\subsection{Simulation setup}
\label{sec:approach:simulations}

\textit{Code and units.} We carry out a suite of simulations of tori with small radial extent using the GPU-accelerated, 3D general-relativistic (magneto-)hydrodynamic (GR(M)HD) code \verb|H-AMR| \citep{HAMR}, although here we do not evolve magnetic fields. Our BH is non-rotating -- we initialize the warp (usually driven by BH rotation) by hand in a manner we will describe shortly. We use a spherical polar grid in $(r,\theta,\varphi)$ coordinates that is centered on the BH. The base grid is uniform in $({\rm log}r,\theta,\varphi)$. We employ adaptive mesh refinement (AMR), allowing us to focus our resolution on regions of higher density. Specifically, we add AMR levels whenever $\rho>0.5\rho_{\rm max}$, where $\rho_{\rm max}=1$ is the initial density maximum. We also employ local adaptive timestepping, which allows different blocks (even at the same AMR level) to evolve at different timesteps, increasing the accuracy and speed of our simulations \citep{koushik_2019}. Our simulations use scale-free units, $G=M=c=1$, such that the BH gravitational radius $r_{\rm g}=GM/c^2=1$, where $M$ is the BH mass. 

\textit{Initial torus.} We initialize axisymmetric, isentropic tori in the manner of \cite{chakrabarti_1985}, except with a novel specific angular momentum profile that is appropriate for both narrow rings and disks (Appendix \ref{app:gr_equilibrium}),
\begin{equation}
    l = l_{\rm c}(\lambda)\left[1 - \epsilon_{\rm torus}^2\left(\frac{\lambda-\lambda_0}{\lambda_0}\right)\right]
    \label{eq:amom_dist}
\end{equation}
Here, $\lambda$ is the von Zeipel radius, which is asymptotically the cylindrical radius, and $l_{\rm c}(\lambda)$ is the angular momentum of circular orbits at the midplane. The torus pressure maximum is located at $\lambda_0$, which equals spherical radius $r_0$ at $\theta=\pi/2$. The aspect ratio of the torus is $\epsilon_{\rm torus}$.
We are also free to choose the inner radius of the torus, $r_{\rm in}$. The fluid-frame gas pressure ($p$) and density ($\rho$) are related by a polytropic equation of state $p=K\rho^\gamma$ where $\gamma$ is the adiabatic index and $K$ is the entropy constant. When $r_0 \gg r_{\rm in}$, the solution is radially extended and disk-like,  with $\epsilon_{\rm torus} \sim H/r$. When $(r_0-r_{\rm in})/r_0\ll 1$, the solution is radially narrow and ring-like, with $\epsilon_{\rm torus} \sim \epsilon$, where $\epsilon$ is the aspect ratio of the rings described in Section \ref{sec:ring}. The main parameter we vary is the initial warp amplitude, $\psi_0$. We initialize warps by performing a radially-dependent rotation of the torus along the (global) $x$-axis by angle $\psi_0 (r-r_0)$. 


\begin{table*}[]
\begin{tabular}{|l|l|l|l|l|l|l|l|}
\hline
Simulation     & $\epsilon_{\rm torus}$ & $\psi_0$ & Base Resolution ($N_r\times N_\theta\times N_\varphi$) & \# AMR levels & $\Delta \theta$ & Initial Condition \\ \hline
L\_e0.1p0.05   & 0.1        & 0.05     & $2048\times256\times768$                               & 0             & $\pi/38$        & Linear "$+$" Tilting Mode                     \\ \hline
W\_e0.01p0.025 & 0.01       & 0.025    & $1536\times1536\times256$                              & 2             & $\pi/5$         & Warp Only                     \\ \hline
W\_e0.01p0.05  & 0.01       & 0.05     & $1536\times1536\times256$                              & 2             & $\pi/5$         & Warp Only                     \\ \hline
W\_e0.01p0.1   & 0.01       & 0.1      & $1536\times1536\times256$                              & 2             & $\pi/5$         & Warp Only                     \\ \hline
W\_e0.01p0.2   & 0.01       & 0.2      & $1536\times1536\times256$                              & 2             & $\pi/5$         & Warp Only                     \\ \hline
W\_e0.01p0.4   & 0.01       & 0.4      & $1536\times1536\times256$                              & 2             & $\pi/5$         & Warp Only                     \\ \hline
WH\_e0.01p0.4  & 0.01       & 0.4      & $1536\times3072\times256$                              & 2             & $\pi/5$         & Warp Only                     \\ \hline
\end{tabular}
\caption{Table of all simulations performed. The simulation naming convention begins with ``L/W/WH'', which are shorthands for ``linear mode'', ``warped'', and ``warped high resolution'', and then the torus thickness and warp amplitude are listed in the name as well.  $\epsilon_{\rm torus}$ is the thickness of the torus solution; $\psi_0$ is the initial warp amplitude; ``Base Resolution'' is the resolution of the grid before AMR levels are applied; ``\# AMR levels'' is the number of AMR levels used, where density is used as the refinement criterion; $\Delta \theta$ is the extent of the computational domain in $\theta$; ``Initial Condition''
indicates how we initialized each simulation, which is either with a linear tilting mode (Eq.~\ref{eq:app:warping_eigenvectors}, which is the generalization of Eq.~\ref{eq:linear_warping_eigenvectors} to arbitrary magnitudes of $\delta/\epsilon$), which includes warping and sloshing motions, or with only a warp.}
\label{table:sims}
\end{table*}

\textit{Grid extent and boundary conditions.} Table \ref{table:sims} shows our full list of simulations and the parameter choices for each. In all cases, we have chosen a $\gamma=5/3$, adiabatic equation of state, but note that entropy may still increase via truncation error or shocks. The torus pressure maximum is at $r_0=50\,r_{\rm g}$ and the torus inner radius is $r_{\rm in}=45\,r_{\rm g}$, such that $\Delta r/r_0=0.1$.  Due to the limited radial and vertical extent of our tori, we can use a restricted computational domain, just large enough to fit the torus. So, while in all cases our grid extends from $\varphi=0$ to $2\pi$ in azimuth, the radial and vertical extents are small: we place the inner and outer radial boundaries at $r=35\,r_{\rm g}$ and $65\,r_{\rm g}$, respectively, and the polar boundaries at $\theta=\pi/2\pm\Delta\theta/2$. The resulting polar extent (which we typically take to be $\Delta\theta=\pi/5$) is much larger than the dimensionless scale height at the pressure maximum of the torus,
\begin{equation}
    h \equiv H/r \approx 10^{-3}\left(\frac{\epsilon_{\rm torus}}{10^{-2}}\right)\left(\frac{\Delta r/r_0}{10^{-1}}\right),
    \label{eq:dimensionless_scale_height}
\end{equation}
for all of our simulations. This accomodates strong warps that can launch material to high latitudes. The limited extent of our computational domain keeps our time step large and allows us to achieve high resolutions with relatively small computational expense. We use periodic boundary conditions in the azimuthal direction and outflow boundary conditions in the radial and polar directions. We terminate each simulation after about twenty orbits, which proved long enough to capture the response of the tori to their warps. 

\textit{Grid resolution.} We choose the resolution such that there are at least $\sim10$ cells per torus scale height and that the $r$ and $\theta$ cell extents are roughly the same size (see Table \ref{table:sims}). The $\varphi$ cell extents are much larger, which is acceptable because the characteristic azimuthal length scale is the torus circumference, which is much longer than the scale height. We will study some simulations where the scale height undergoes strong oscillations, and in these cases the scale height minima are still resolved by $\sim5$ resolution elements. 

\subsection{Simulating linear modes}
\label{sec:approach:validation}

We want to use our simulations to determine the extent to which we can use the ring theory. We first show that the two descriptions agree in simple cases by initializing a torus with aspect ratio $\epsilon_{\rm torus}=0.1$ with a linear tilting mode. This is thicker than most tori we will consider later, but
it is useful as it makes the precession frequency in the global frame\footnote{The relationship in Equation \ref{eq:precession_frequency} is approximate because it neglects relativistic effects. This approximation is only made for the clarity of our presentation and is not used in our calculations.},
\begin{equation}
    \omega_{\rm p} \approx \Omega-\omega_\psi
\label{eq:precession_frequency}
\end{equation}
higher (see Eq.~\ref{eq:apprx_bending_eigenvalues}). Here, $\omega_{\rm p}<0$ indicates retrograde precession and $\omega_{\rm p}>0$ indicates prograde precession. 
As a result,
it requires a shorter run to simulate for a few precession times ($2\pi/\omega_{\rm p}$), and simulating multiple precession times is a better test of the ring theory.
We set $\psi_0=0.05$, which is just small enough such that the warp behaves linearly. We initialize horizontal velocities that satisfy the eigenvector relationship (Eq.~\ref{eq:app:warping_eigenvectors}), which is necessary to simulate a pure ``$+$'' tilting mode that we can easily compare with the ring theory. 
To do this we must connect $\epsilon_{\rm torus}$ to the ring aspect ratio $\epsilon$. These two quantities are of the same order, and we find that setting $\epsilon=1.37\epsilon_{\rm torus}$ results in a good quantitative match between the simulated and predicted ratio of the eigenvector amplitudes, $|\psi|/|\sigma|$. 
In order to input the eigenvectors from the ring theory into the simulation, we must include their $\varphi$-dependence, which is accomplished by appending ${\rm exp}(i\varphi)$ to $\sigma$ in Equation \ref{eq:app:warping_eigenvectors}. 
The ``$+$'' tilting mode that we initialize undergoes retrograde precession (Eq.~\ref{eq:precession_frequency}). 

We show the result in Figure \ref{fig:gentle_ring}. On the left, we depict four panels consecutively in time. Each depicts a density map of the torus in the $r$-$\theta$ plane. Each slice is depicted in the rotating reference frame, such that snapshots follow the orbital motion of the same parcel of gas. Note that the torus is not exactly symmetric -- it stretches from $x/r\approx-0.1$ to $0.12$ -- and so differs modestly from the symmetric assumption of the ring theory. As the torus orbits, it gently rocks back and forth by an 
angle $\approx\psi_0$. 

\begin{figure*}
    \centering
    \includegraphics[width=\textwidth]{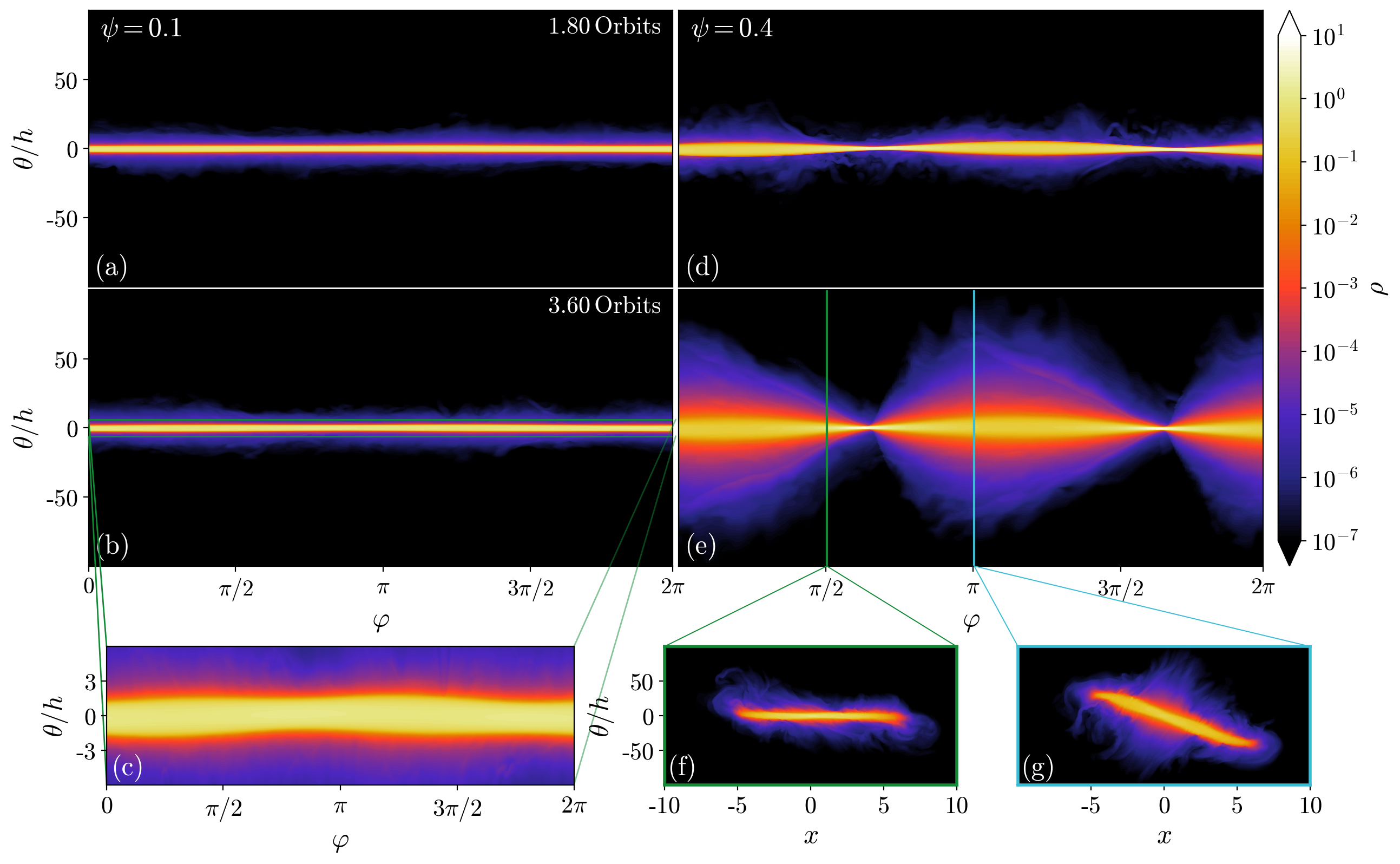}
    \caption{Sequence of $\varphi-\theta$ torus density slices for simulations W\_e0.01p0.1 ($\psi_0=0.1$) and W\_e0.01p0.4 ($\psi_0=0.4$).
    This figure shows that when $\psi_0$ becomes large enough, the scale height oscillations become extreme. \textbf{Panels a-b.} At $\psi_0=0.1$, the scale height barely oscillates. \textbf{Panel c.} We zoom in vertically on the scale height to show its small oscillations. \textbf{Panels d-e.} At $\psi_0=0.4$, the warp is above a critical threshold such that it forces strong vertical oscillations -- this is the ``bouncing'' regime. The gas oscillates vertically twice an orbit, with the scale height reaching amplitudes on the order of the equilibrium scale height, while lower density gas ($\sim10^{-3}-10^{-4}\times$ the midplane value, blue) is displaced to latitudes two orders of magnitude higher than the equilibrium scale height. 
    \textbf{Panels f-g.} We show $x-\theta$ cross-sections of the flow at the same time as panel e, 
    where $x=r-r_0$ and the $\varphi$ coordinates of these panels are marked by colored-coded vertical lines in panel e.}
    \label{fig:bouncing_onset}
\end{figure*}

In the right three panels we compare the simulated results to the ring equations (Eqs. \ref{eq:fa:1}-\ref{eq:fa:10}). We label the panels ``warping'' ($\psi$), ``sloshing'' ($\sigma$) and ``breathing'' ($\eta$). Whereas the ring equations evolve these quantities, we must measure them in the simulations. To do so, we project our four-velocities onto the orthonormal tetrad carried by an observer orbiting circularly at $r_0$ with frequency $\Omega$ (Eq.~\ref{eq:tetrad}). This is the general-relativistic version of transforming to the rotating reference frame and it results in a set of local velocities ${\rm \textit{v}}_i =({\rm \textit{v}}_x,{\rm \textit{v}}_y,{\rm \textit{v}}_z)$. Using these velocities, we solve for the flow matrix $A_{ij}$ by fitting the density-weighted velocities to the right hand side of Equation \ref{eq:flow_matrix} over the extent of the ring at each time and azimuth. We then convert to the local time coordinate of the relativistic shearing sheet (Eq.~\ref{eq:lc_coord:t}). 

The result shows that $\psi$, $\sigma$ and $\eta$ behave the same in the simulation as in the ring theory. While $\psi$ and $\sigma$ evolve as a pure tilting mode, $\eta$ has a beat frequency. This is because there are two $\eta$ modes present. The initial condition excites $\eta$'s natural mode, which oscillates at frequency $\omega_\eta$. The warp drives the forced mode, which oscillates at frequency $2\omega_\psi$ (Eq.~\ref{eq:ql_superposition}). 

We consider the match shown in Fig.~\ref{fig:gentle_ring} excellent. 
The ring theory is local and has restricted degrees of freedom, the simulations are global and relativistic, and yet the two approaches largely agree. In the following sections, we will push this correspondence until it breaks down. 


\begin{figure}
    \centering
    \includegraphics[width=\textwidth]{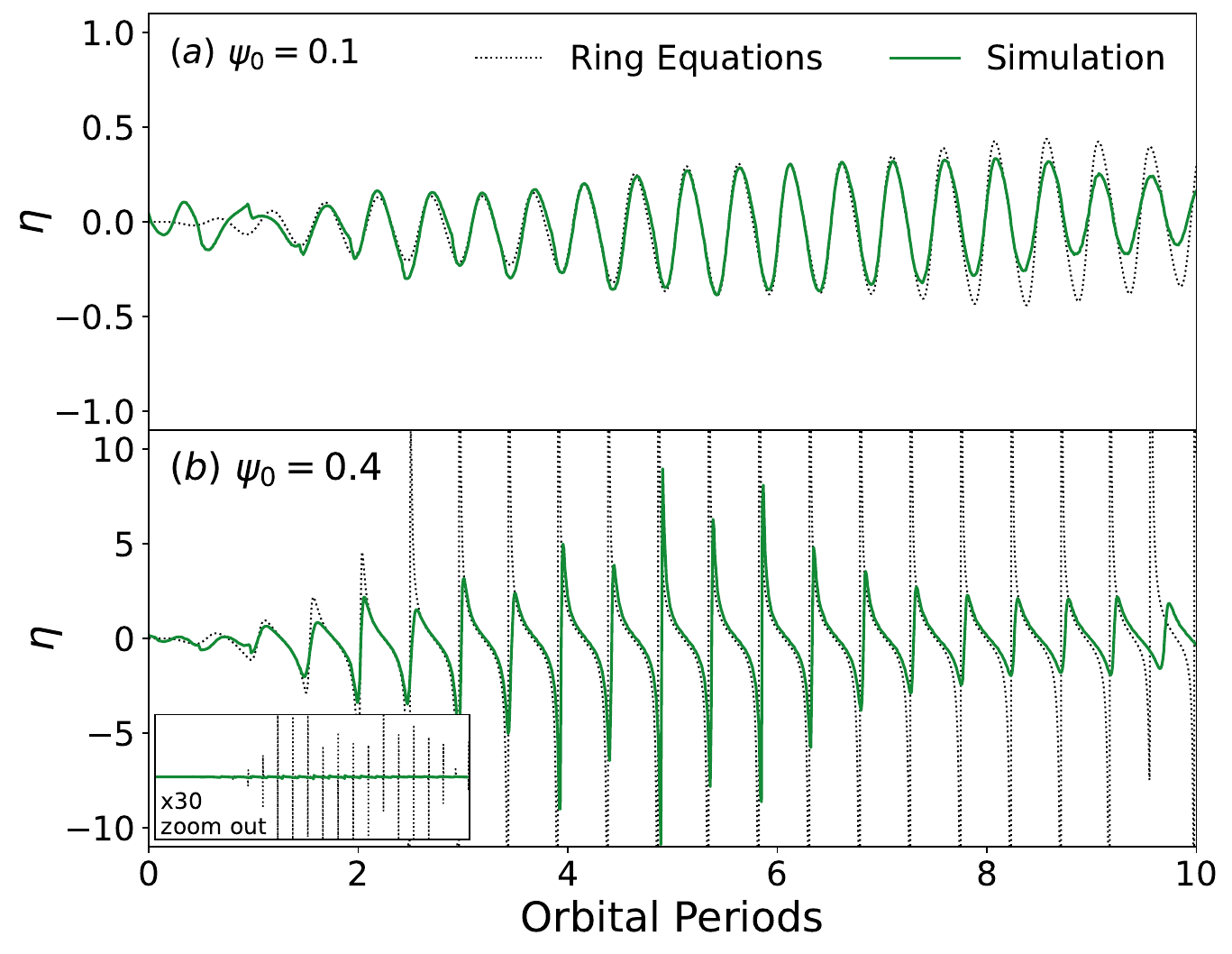}
    \caption{Comparison of simulated $\eta(t)$ to ring theory $\eta(t)$ for simulations W\_e0.01p0.1 ($\psi_0=0.1$) and W\_e0.01p0.4 ($\psi_0=0.4$). \textbf{Panel (a).} At $\psi_0=0.1$, the ring does not bounce, and the ring theory and simulated $\eta$ largely agree. \textbf{Panel (b).} At $\psi_0=0.4$, the evolution of $\eta$ is much more dramatic, as the ring undergoes bouncing. While the two approaches agree almost everywhere, the ring theory predicts much higher peak values of $\eta$. Note that the vertical axis in panel (b) is ten times larger than panel (a). We also show an inset panel with the vertical axis extending an additional factor of thirty larger.
    }
    \label{fig:a33_evo}
\end{figure}

\section{Bouncing Regime}
\label{sec:bouncing_regime}

\subsection{An empirical description}
\label{sec:bouncing_regime:empirical}

When \cite{fairbairn_ogilvie_2021b} (``FO21b'') initialized rings above a critical warp amplitude, the rings underwent extreme scale height oscillations. They referred to this as the ``bouncing'' regime. K23 reported an extremely high-resolution, global GRMHD simulation of a thin accretion disk that was highly tilted with respect to the BH spin axis. This simulation featured warps with similarly extreme scale height oscillations that dramatically affected the accretion process. These two results suggest that the bouncing regime is both robust and dynamically important, meriting further exploration.

Figure \ref{fig:bouncing_onset} depicts the time evolution of density for a pair of tori simulations over the course of several orbits. In the first column, we show a torus with $\psi_0=0.1$, and in the right, a torus with $\psi_0=0.4$. Both tori have $\epsilon_{\rm torus}=0.01$. In both columns, the horizontal axis is periodic in $\varphi$ and the vertical axis shows $\theta$ normalized to $h$ in the equilibrium state. There is a clear difference in evolution between the two tori: at $\psi_0=0.1$, the torus is nearly flat and the density contours oscillate only mildly from their initial height. Yet, when $\psi_0=0.4$, an extremely strong scale height oscillation twice an orbit develops within only a few orbits. This indicates that the $\psi_0=0.4$ torus is in the bouncing regime. In Fig.~\ref{fig:bouncing_onset}(f)-(g), we also show $r-\theta$ cross-sections of the ring depicted in Fig.~\ref{fig:bouncing_onset}(e), where there are two vertical color-coded lines that correspond to these cross-sections. This figure demonstrates that there is a dramatic shift in behavior of the tori between $\psi_0=0.1$ and $\psi_0=0.4$. 

We compare the $\psi_0=0.1$ and $\psi_0=0.4$ simulations further in Figure \ref{fig:a33_evo}, where we show the simulated $\eta$ as a function of time (green). The values are presented in the rotating reference frame to compare with the ring theory $\eta$ (black). In Fig.~\ref{fig:a33_evo}(a), $\psi_0=0.1$, and the simulation $\eta$ mostly matches the ring theory $\eta$. In Fig.~\ref{fig:a33_evo}(b), the shape and phase of $\eta$ is similar between the ring theory and the simulations, but the peak amplitudes disagree. The simulations produce peak values of $\eta$ in the range $1-10$, yet the ring theory predicts that $\eta$ peaks at values that are orders of magnitude larger. This is clear in the inset of Fig.~\ref{fig:a33_evo}(b), where we show the same profiles except with a vertical axis that is thirty times larger\footnote{We found that, in the bouncing regime, the peak values of the ring theory $\eta$ scale roughly inversely with the integrator time-step and did not converge.}. These results indicate that in three-dimensional simulations of the bouncing regime, there is some mechanism preventing the scale height oscillations from reaching the heights predicted by the ring theory. 

\subsection{The onset of bouncing}
\label{sec:bouncing_regime:leadup}

\begin{figure}
    \centering
    \includegraphics[width=\textwidth]{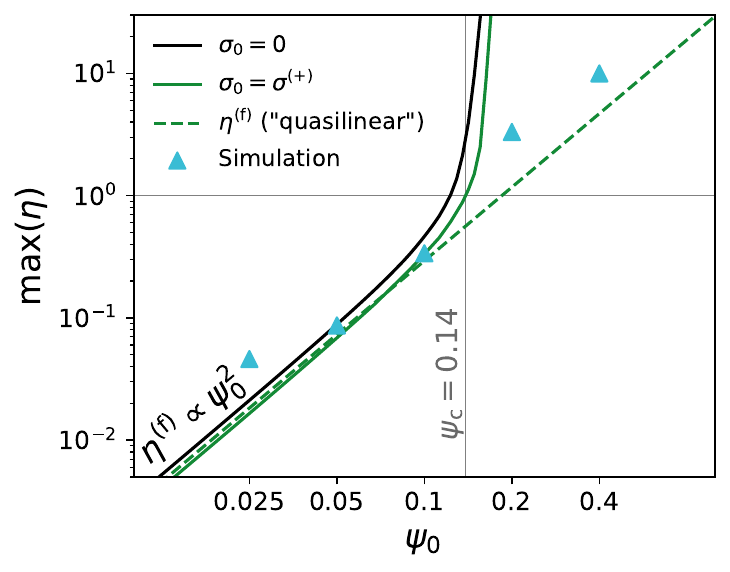}
    \caption{Comparison of maximum $\eta$ (${\rm max}(\eta)$) between the ``quasi-linear'' ring theory, the fully nonlinear ring theory and simulations as a function of initial warp amplitude $\psi_0$. When $\eta$ exceeds unity, tori begin bouncing. We show $\eta^{\rm (f)}$, which is our quasilinear prediction for ${\rm max}(\eta)$ in the small $\psi_0$ regime  (Eq.~\ref{eq:forced_breathing}, green dashed). 
    We integrate the ring equations for zero initial slosh ($\sigma_0=0$, black solid) and a tilting mode ($\sigma_0=\sigma^{(+)}$, green solid). We also show ${\rm max}(\eta)$ for simulations W\_e0.01p0.025 through W\_e0.01p0.4 (blue triangles), which is calculated by taking the maximum value of $\eta$ over the course of each simulation.} 
    \label{fig:bouncing1d}
\end{figure}

We determine the onset of bouncing by examining where linear theory breaks down.  In Figure \ref{fig:bouncing1d}, we measure the maximum value of $\eta$ for separate integrations of the ring equations (solid lines) as a function of $\psi_0$. We initialize the ring equations with either a vertical tilting mode  (``$\sigma_0=\sigma^{(+)}$'', green) or a warp with zero initial slosh (``$\sigma_0=0$'', black). Both behave similarly. When ${\rm max}(\eta)\gtrsim1$, the breathing becomes strongly nonlinear and the rings enter the bouncing regime, as indicated by the steep growth of ${\rm max}(\eta)$ with $\psi_0$. We have also plotted the maximum value of $\eta$ recorded during the runtime of our three-dimensional simulations (blue triangles). As in Figure \ref{fig:a33_evo}, the simulated $\eta$ do not achieve the extreme amplitudes predicted by the ring theory. We will explain this difference in Section \ref{sec:bouncing_regime:nozzle_shocks}. 

We predict the onset of bouncing by returning to the coupling of tilting and breathing modes described in Section \ref{sec:rings:linear:quasi}. Tilting modes force breathing modes $\eta^{\rm (f)}\propto|\psi\sigma|$ at second order (Equation \ref{eq:forced_breathing}). We regard these forced breathing modes as ``quasilinear'' since they are formally linear despite resulting from a second order coupling.
We can expand $\eta^{\rm (f)}$ by inputting the linear tilting eigenvectors for $\psi$ and $\sigma$ (``$+$'' solution in Equation \ref{eq:app:warping_eigenvectors}, which is the generalization of Eq.~\ref{eq:linear_warping_eigenvectors} to arbitrary magnitudes of $\delta/\epsilon$),
\color{black}
\begin{equation}
        |\psi\sigma| = \frac{2|\psi|^2}{\delta+\sqrt{\delta^2 + 4\epsilon^2}}
        \label{eq:forcing_nolimit}
\end{equation}
In the Keplerian ($\delta\rightarrow0$) limit this expression becomes,
\begin{equation}
    \underset{\epsilon\gg\delta}{\rm lim}|\psi\sigma| = \frac{|\psi|^2}{\epsilon}
    \label{eq:forcing_keplerian}
\end{equation}
 This is the same forcing found in FO21b. We care about the non-Keplerian limit, where
\begin{equation}
    \underset{\delta\gg\epsilon}{\rm lim}\,\,|\psi\sigma| = \frac{|\psi|^2}{\delta}
\label{eq:forcing_+}
\end{equation}

In either case, $\eta^{\rm (f)}\propto\psi^2$ (Eq.~\ref{eq:forced_breathing}). We depict $\eta^{\rm (f)}$ (green dashed line) in Figure \ref{fig:bouncing1d}. It agrees with the ring theory predictions\footnote{We also show this for a larger parameter space in Figure \ref{fig:bouncing1d_app} in the Appendix.} and (roughly) with the simulations up to about $\psi_0\lesssim0.1$. 

The ring theory suggests that bouncing occurs when $\psi_0\gtrsim0.14$, which is shown in Fig.~\ref{fig:bouncing1d} by the steepening of the green and black curves. This happens when ${\rm max}(\eta)\approx1$, which we will explain physically in the following section. We estimate $\psi_{\rm c}$ for other values of $\delta$, $\epsilon$ and $\gamma$ by using quasilinear theory, where we set\footnote{In general, the critical $\eta^{\rm (f)}$ depends on $\gamma$, $\delta$, and $\epsilon$, as can be seen in Figure \ref{fig:bouncing1d_app} in the Appendix. Since this dependence is weak, it is sufficient to set $\eta^{\rm (f)}$ to $0.57$.} $\eta^{\rm (f)}=0.57$. We use this value because it corresponds to the value of $\psi_0$ at which ${\rm max}(\eta)=1$ for the green curve in Fig.~\ref{fig:bouncing1d}. By inputting this estimate into Eq.~\ref{eq:forced_breathing}, using Eq.~\ref{eq:forcing_nolimit}, and solving for $|\psi|$, we find

\begin{equation}
    \psi_{\rm c} \approx 0.75\sqrt{\left(\frac{3-\gamma}{6+2\gamma - 4\delta/3(\gamma-1)}\right)\left(\delta + \sqrt{\delta^2 + 4\epsilon^2}\right)},
\label{eq:psi_crit}
\end{equation}
where we have labeled the warp amplitude $\psi_{\rm c}$ to indicate that it is the critical warp amplitude above which rings bounce. In the Keplerian ($\delta\rightarrow0$) limit, assuming $\gamma=5/3$, 
\begin{equation}
    \psi_{\rm c} \approx 0.04\left(\frac{\epsilon}{0.01}\right)^{1/2}
\label{eq:psi_crit_keplerian}
\end{equation}
This is consistent to order unity with FO21b, who found $\psi_{\rm c}\approx0.06\left(\frac{\epsilon}{0.01}\right)^{1/2}$.
In the non-Keplerian limit ($\delta\gg\epsilon$), the critical warp amplitude is,
\begin{equation}
    \psi_{\rm c} \approx 0.14\left(\frac{\delta}{0.12}\right)^{1/2}
\label{eq:psi_crit_nonkeplerian}
\end{equation}
Figure \ref{fig:bouncing1d} shows $\psi_{\rm c}$ with a vertical line, which predicts the $\psi_0$ where bouncing sets in up to an order-unity prefactor that depends weakly on $\sigma_0$.

\subsection{The activation of resonant bouncing}
\label{sec:bouncing_regime:resonance}

In Section \ref{sec:bouncing_regime:empirical}, we empirically showed that both the ring theory and the three-dimensional simulations enter the bouncing regime when the warp amplitude is large enough. In Section \ref{sec:bouncing_regime:leadup}, we derived the critical warp amplitude above which bouncing occurs, which corresponds to $|\eta|\approx 1$. But what exactly causes the dramatic rise in $|\eta|$ once $|\psi|>\psi_{\rm c}$?

As described above, the breathing mode is driven quasilinearly by the tilting modes at frequency $2\omega_\psi\approx2\nu$ (Eq.~\ref{eq:ql_superposition}). At small $\eta$, this forcing frequency is larger than the natural frequency of the breathing mode, $\omega_\eta\approx\nu\sqrt{1+\gamma}$ (Eq.~\ref{eq:linear_breathing_eigenvalues}), by an order unity factor. When $\eta$ increases, $\omega_\eta$ increases, until it is close to $2\omega_\psi$. This drives a resonance wherein the warp drives $\eta$ to extremely large amplitudes.

We can provide a more physical picture of the increase of $\omega_\eta$
by expanding upon FO21b's description. Small scale height oscillations are described by linear breathing modes which oscillate at frequency $\omega_\eta\approx\nu\sqrt{1+\gamma}\approx
1.63\nu$ for $\gamma=5/3$. We can take $\eta\approx\delta H/H$. When $\eta\approx1$, $\delta H\approx H$, and the ring is strongly compressed. This activates impulsive pressure forces at the scale height minima, launching gas to high latitudes. At high latitudes, the pressure force is weak, and gas is pulled down by gravity. As described by FO21b, this is analogous to a ball bouncing off a rigid table. Since the gas spends most of its time between bounces, it will try and oscillate at frequency $\nu$. However, half-way through the vertical oscillation, gas parcels land at a scale height minimum and their motion is impulsively reversed. One ``bounce'' then lasts for period $\approx \pi/\nu$. The impulsive pressure force advances the phase somewhat, which shortens the period, but this is a higher order effect which we ignore here. So, when a scale height oscillation is large enough, it evolves with period $\approx\pi/\nu$ which corresponds to an angular frequency $\gtrsim2\nu$. Therefore, the natural breathing frequency increases from $\approx1.63\nu$ (for $\gamma=5/3$) to $\approx2\nu$ as one proceeds from small to large $\eta$. Since the quasilinear forcing frequency is $2\omega_\psi\approx2\nu$ as well, the warp resonantly forces breathing at large $\eta$.

\begin{figure}
    \centering
    \includegraphics[width=\textwidth]{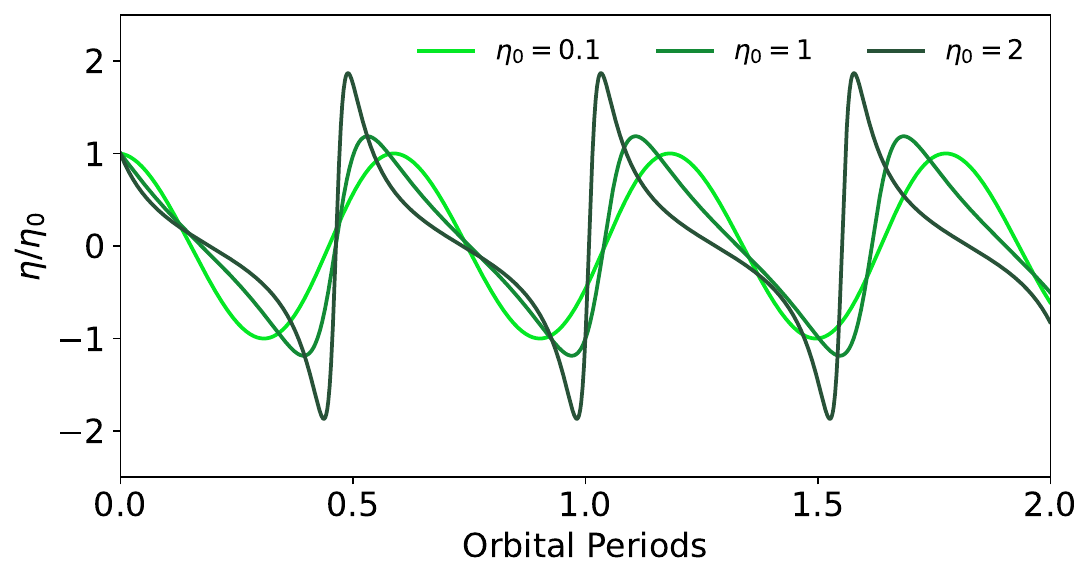}
    \caption{Comparison of normalized ring theory $\eta$ for initial values $\eta_0=0.1$, $1$ and $2$, with $\epsilon=0.01$, $\delta=0.12$ and no warp. The ring theory predicts that pure breathing motions ($\psi=\sigma=0$) have linear frequencies somewhat larger than the orbital frequency. As $\eta$ increases and nonlinear effects become important, these frequencies approach a value near two. We can see
    that as $\eta_0$ increases, the position of the first peak approaches roughly half an orbital period.}
    \label{fig:pure_breathing}
\end{figure}

We demonstrate the increase of $\omega_\eta$ in Figure \ref{fig:pure_breathing}, where we show three integrations of the ring equations with initial breathing amplitudes, $\eta_0=0.1$, $1$ and $2$, and no initial warp. At $\eta_0=0.1$, the breathing motions are linear, and evolve at frequency $\omega_\eta$. As $\eta_0$ increases, nonlinear effects come into play, and this frequency increases to a value that is $\approx2\nu$. Since the warp-induced forcing also has a frequency that is $\approx 2\nu$, it can resonantly couple to the free breathing mode in the $|\eta|>1$ regime. This allows the warp to force the scale height to large values. This resonance is why the ring theory predicts extreme amplifications of $|\eta|$ above $\psi_{\rm c}$, as demonstrated by the solid lines in Fig.~\ref{fig:bouncing1d}.

\subsection{Nozzle Shocks}
\label{sec:bouncing_regime:nozzle_shocks}

We now return to the breathing and bouncing of our simulated tori. In Figure \ref{fig:bouncing1d}, in addition to the ring theory results, we show the maximum breathing amplitude recorded over the course of our simulations (blue triangles). Below $\psi_{\rm c}$, the simulated ${\rm max}(\eta)$ usually agrees with both the ring equations and $\eta^{\rm (f)}$, as expected. The one exception is $\psi_0=0.025$, wherein ${\rm max}(\eta)$ is about twice as high as expected. We found that this maximum occurs within one orbital period and thereafter is well-described by the ring theory, so we attribute this to an early transient\footnote{A similar transient increase in $\eta$ can be seen within the first orbital period of Fig.~\ref{fig:a33_evo}(a), however it is relatively small in this case, since the $\psi_0=0.01$ torus drives a larger $\eta$ than $\psi_0=0.025$.}. We also  find that even above $\psi_{\rm c}$, the simulated ${\rm max}(\eta)$ roughly obeys the $\propto \psi_0^2$ scaling, which is not expected in the bouncing regime. However, $|\eta|$ does not remain high in the $\psi_0>\psi_{\rm c}$ simulations. This may be seen directly for the $\psi_0=0.4$ simulation in Fig.~\ref{fig:a33_evo}(b), where $|\eta|$ peaks at roughly five orbital periods but then begins decreasing in amplitude to values much lower than the ring theory predicts. We attribute this decay to shock dissipation. 

Our goal is to understand why and how shocks occur in the bouncing regime. When tori bounce, gas is forced through scale height minima, where the vertical component of velocity reverses direction. We describe these regions as ``nozzles''. Unlike the more familiar nozzles that occur when fluid is forced through a pipe of small cross-section, here the nozzles occur because gravity squeezes the gas vertically. At the radial center of the torus, the vertical velocity is $v_z\approx\eta H\Omega$ one scale height above the midplane. Since the sound speed is $c_{\rm s}\approx H\Omega$ and in the bouncing regime $|\eta|\gtrsim 1$, the vertical velocity here is supersonic. This naturally forms shocks in the nozzles, which we call ``nozzle shocks'', as in K23. These are akin to the nozzle shocks that occur in tidal disruption events \citep{evans_kochanek_1989}.  

\begin{figure}
    \centering
    \includegraphics[width=\textwidth]{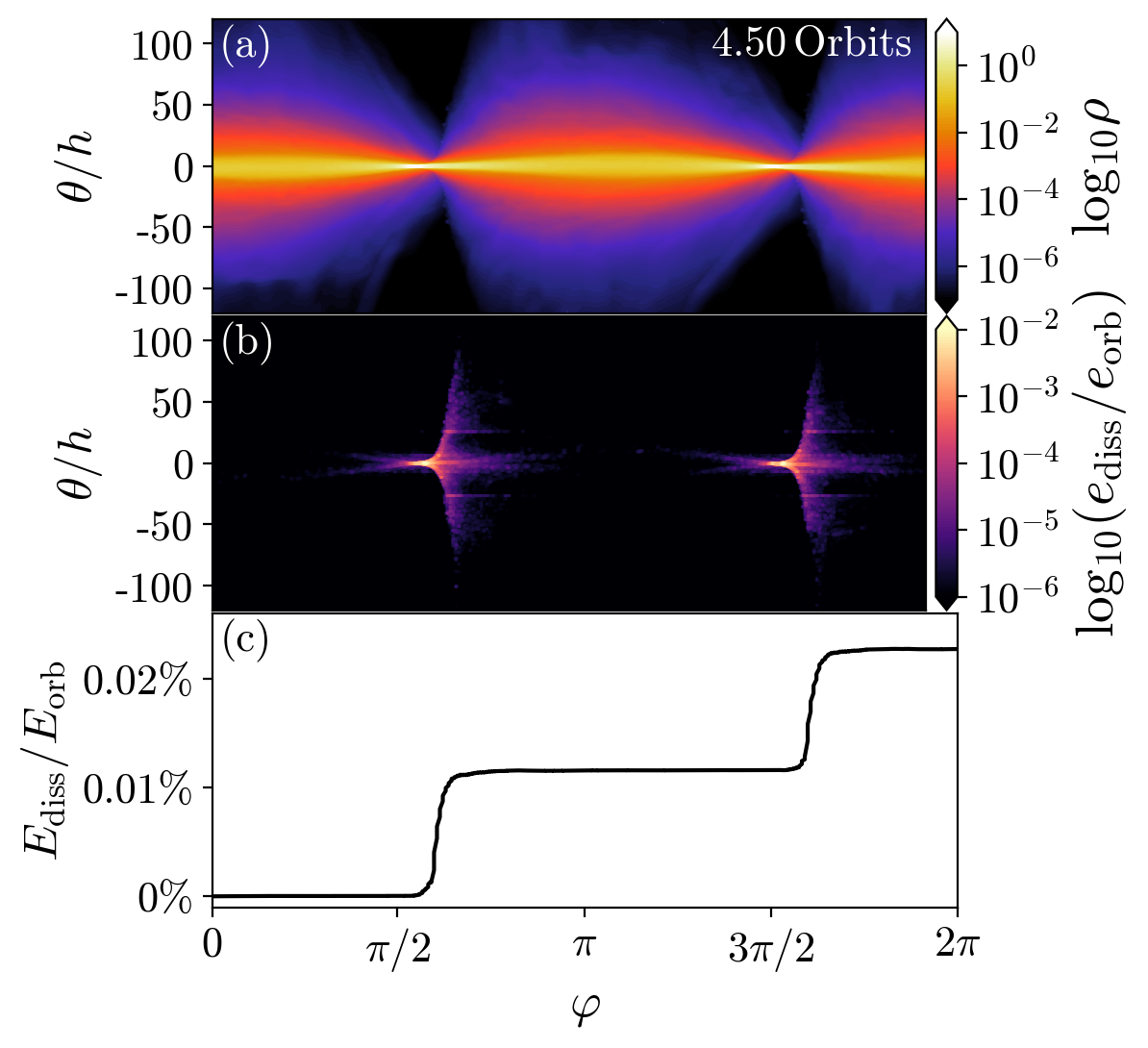}
    \caption{Depiction of shocks in the bouncing regime for simulation WH\_e0.01p0.4.    
    When rings bounce, they excite nozzle shocks twice an orbit that dissipate the orbital energy of the gas. \textbf{Panel a.} We depict a $\varphi-\theta$ slice of density contours after about four orbits, where we can see that the density scale height experiences strong oscillations twice an orbit. \textbf{Panel b.} We depict energy dissipation (Eq.~\ref{eq:dissipation_per_orbit}), normalized to the orbital energy density ($e_{\rm orb}$ in Eq.~\ref{eq:specific_energy_orbit}), at the same time as panel a. We can see that there is strong dissipation where the gas density is compressed, indicating the presence of shocks. \textbf{Panel c.} We depict the cumulative integration of energy dissipation along $\varphi$ (Eq.~\ref{eq:cumulative_dissipation}), normalized to the vertically-integrated orbital energy of the gas ($E_{\rm orb}$ in Eq.~\ref{eq:specific_energy_orbit}). We can see step function like features where the gas shocks, which dissipate $\mathcal{O}(0.01\%)$ of the orbital energy of the gas.}
    \label{fig:nozzle_shock}
\end{figure}

In Figure \ref{fig:nozzle_shock}, we show an example of nozzle shocks in our high-resolution $\psi_0=0.4$ simulation, WH\_e0.01p0.4 after 4.5 orbits. In Fig.~\ref{fig:nozzle_shock}(a), we depict a $\varphi-\theta$ slice of gas density, where it is clear that the torus is in the bouncing regime since the scale height oscillates dramatically twice an orbit. In Fig.~\ref{fig:nozzle_shock}(b), we depict the same slice, except depicting a measure of dissipation in shock fronts. To measure dissipation, we begin by expressing the heating rate per unit volume of a fluid parcel by its change in entropy, 
\begin{equation}
    Q=\rho^\gamma(\gamma-1)^{-1}u^\mu\partial_\mu\kappa_{\rm g},
\label{eq:dissipation_per_time}
\end{equation}
where $\kappa_{\rm g}$ is the specific entropy of the gas. In steady-state, this quantity is zero except where there is dissipation, since specific entropy is conserved along streamlines. While our simulation is not in steady state, this quantity still proves to be an effective measure of dissipation. Although our simulation uses an adiabatic $\gamma=5/3$ equation of state, dissipation may still occur numerically either via truncation error or via our shock capturing scheme. We also define
\begin{equation}
e_{\rm diss}= Q/\Omega,    
\label{eq:dissipation_per_orbit}
\end{equation}
which is an estimate of the energy dissipated on an orbital timescale. We plot this in Fig.~\ref{fig:nozzle_shock}(b), where we have normalized it to the orbital energy density\footnote{The $+1$ in Eq.~\ref{eq:specific_energy_orbit} removes the rest-mass energy contribution from $u_t$.} $\rho(u_t+1)$ averaged at $r_0$,
\begin{equation}
\begin{aligned}
    e_{\rm orb}&=\frac{E_{\rm orb}}{\int_0^{2\pi} \sqrt{g_{\varphi\varphi}}d\varphi\int_{\theta_{\rm min}}^{\theta_{\rm max}}\sqrt{g_{\theta\theta}}d\theta }\bigg|_{r=r_0}\\
    &=\frac{\int_0^{2\pi} \sqrt{g_{\varphi\varphi}}d\varphi\int_{\theta_{\rm min}}^{\theta_{\rm max}}\sqrt{g_{\theta\theta}}d\theta \rho(u_t+1)}{\int_0^{2\pi} \sqrt{g_{\varphi\varphi}}d\varphi\int_{\theta_{\rm min}}^{\theta_{\rm max}}\sqrt{g_{\theta\theta}}d\theta }\bigg|_{r=r_0},
\label{eq:specific_energy_orbit}
\end{aligned}
\end{equation}
where $E_{\rm orb}$ is the orbital energy per unit radius of the ring and $u_t$ is the time component of the covariant four-velocity. We can see that at the nozzles, the dissipation peaks. The dissipation is also asymmetric, with more vertically extended dissipation to the right of the nozzle. This is because the flow moves from the left to right and high latitude streamlines overshoot the nozzle. In Fig.~\ref{fig:nozzle_shock}(c), we again show shock dissipation at $r_0$, except integrated vertically and cumulatively integrated along the orbit,
\begin{equation}
    E_{\rm diss}=\int_0^\varphi\sqrt{g_{\varphi'\varphi'}}d\varphi'\int_{\theta_{\rm min}}^{\theta_{\rm max}}\sqrt{g_{\theta\theta}}d\theta e_{\rm diss}\bigg|_{r=r_0},
\label{eq:cumulative_dissipation}
\end{equation}
which we have normalized to $E_{\rm orb}$. Were there no dissipation, $E_{\rm diss}(\varphi)$ would be zero everywhere. However, we see two clear step function like features at the positions of the nozzles. In each, $\mathcal{O}(0.01\%)$ of the orbital energy of the ring is dissipated. This value can be understood as follows. The non-relativistic specific orbital energy, which is accurate at $50\,r_{\rm g}$, is $-GM/2r$ for circular orbits. The specific vertical kinetic energy at the torus center is $\frac{1}{2}v_z^2$. If all of this energy is dissipated, and we assume most of the fluid has Keplerian velocity $v_{\rm k}^2=GM/r$, then the fraction of energy dissipated per nozzle shock is $v_z^2/v_{\rm k}^2$. In the ring formalism, $v_z\approx \eta H\Omega$. Taking $v_{\rm k}=r\Omega$, the fraction of orbital energy dissipated per nozzle shock is approximately,
\begin{equation}
    \frac{E_{\rm diss}}{E_{\rm orb}} \approx h^2|\eta|^2 = 10^{-4}\left(\frac{h}{10^{-3}}\right)^2\left(\frac{|\eta|}{10}\right)^2,
    \label{eq:ediss_est}
\end{equation}
This is consistent with both our findings and the findings of K23, who found fractional dissipation rates that were $\mathcal{O}(1\%)$ but in a disk of aspect ratio $h=0.02$.

The snapshot depicted in Figure \ref{fig:nozzle_shock} was chosen to be one of the stronger examples of nozzle shocks, where $|\eta|$ is consistent with the maximum value of ${\rm max}(\eta)\approx10$ shown in Figure \ref{fig:bouncing1d}. Since the dissipation is strong, in steady disks $\eta$ may be limited to values that are not too far above unity. It is interesting that Equation \ref{eq:ediss_est} suggests that each nozzle shock dissipates energy roughly equal to the thermal energy ($E_{\rm th}\approx h^2 E_{\rm orb}$). For the disk to remain thin, it must cool this excess heat faster than an orbital timescale, or else the scale height will puff up on a timescale,
\begin{equation}
    t_{\rm puff} \equiv h/\dot{h} \approx \frac{1}{2|\eta|\Omega},
    \label{eq:tau_puff}
\end{equation}
where the factor of $2$ results from having two nozzle shocks per orbit. Indeed, we found that $h$ increased rapidly after nozzle shocks set in (not plotted). Bouncing may be more difficult to achieve if the scale height grows, as when $h$ is large the critical warp amplitude to enter bouncing is also large (see Equation \ref{eq:psi_crit}, where we exchange $\epsilon$ for $h$). However, empirically, the only simulation of a strongly warped general-relativistic thin disk with explicit radiation \citep{liska_kaaz_2023} continues to bounce in the presence of nozzle shocks without puffing up significantly. Furthermore, the possibility of nozzle shocks in thicker, inefficiently cooling disks is reinforced by 
earlier work on warped thick accretion disks that also reported shocks \citep{fragile_blaes_2008,white_2019}\footnote{\cite{fragile_blaes_2008} observed similar shock features in thick accretion disks, which they referred to as ``standing shocks'' -- these and the nozzle shocks studied here may be one and the same.}. Still, the rapid heating suggested by Equation \ref{eq:tau_puff} merits further investigation into the thermodynamics of strongly warped disks.

\begin{figure}
    \centering
    \includegraphics[width=\textwidth]{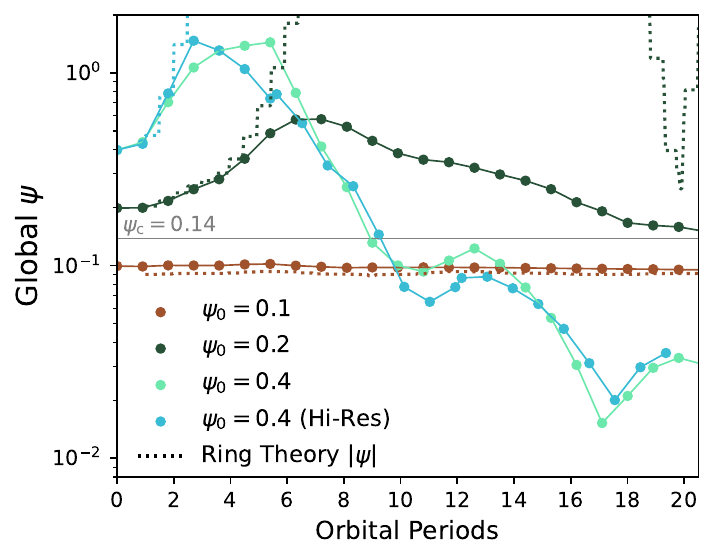}
    \caption{Decay of the simulated warp as a function of time. Global $|\psi|$ is measured by taking the maximum value of the local ($\varphi$-dependent) $\psi$ over the $2\pi$ extent of the simulated tori at a given time. Tori with initial warp amplitude, $\psi_0$, greater than the critical warp amplitude, $\psi_{\rm c}$ (Eq.~\ref{eq:psi_crit_nonkeplerian}), dissipate their warp in $10-20$ orbital periods. We also compare to the non-dissipative ring theory by showing the amplitude of the ring theory $\psi$ (dotted lines) for each $\psi_0$. These amplitudes are calculated by taking the maximum value of $|\psi|$ every orbit. The ring theory $|\psi|$ reaches much higher amplitudes than the simulated $|\psi|$ and never dips below $\psi_0$.}
    \label{fig:warpevo}
\end{figure}

Dissipation in nozzle shocks also damps the warp. We depict this in Figure \ref{fig:warpevo}, where we show $|\psi|$ for the simulations. We measure $\psi$ as we did $\eta$, which is by fitting the simulated velocities to the linear flow field assumed by the ring theory (Eq.~\ref{eq:flow_matrix}) as a function of time and $\varphi$. Here, our ``global'' $\psi$ is defined by taking the maximum value of $\psi(\varphi,t)$ over the $2\pi$ extent of the tori at each time. We have marked the dividing line for bouncing, $\psi_{\rm c}=0.14$ (Eq.~\ref{eq:psi_crit}). Below $\psi_{\rm c}$, our run with $\psi_0=0.1$ shows a steady warp amplitude across the simulation runtime, indicating that there is no dissipation of the warp. However, the runs above this line rapidly dissipate their warp. At $\psi_0=0.2$, the warp amplitude decays to $\psi_{\rm c}$ after about twenty orbits. At $\psi_0=0.4$, for both standard and higher resolution runs, the warp amplitude decays to $\psi_{\rm c}$ at about ten orbits and then continues to decay further. These results suggest that nozzle shocks dissipate warps on the order of ten orbital periods. The warp in the $\psi_0=0.4$ run decays below $\psi_{\rm c}$ because the large values of $\eta$ have not yet decayed sufficiently, as seen at $10$ orbital periods in Fig.~\ref{fig:a33_evo}(b). 

We can confirm that the warp is decaying in our simulations where $\psi_0>\psi_{\rm c}$ by comparing to the non-dissipative ring theory $|\psi|$, which we show with dotted lines in Fig.~\ref{fig:a33_evo}. The ring theory $|\psi|$ is calculated by integrating the ring equations for each $\psi_0$ and taking the rolling maximum of $|\psi|$ every orbital period. At early times, the ring theory $|\psi|$ shows the same early rise as the simulated $|\psi|$, but continues to grow to very large amplitudes. We attribute this growth to the coupling of $\psi$ with the radial breathing of the rings (e.g., Sec.~\ref{sec:rings:linear:breathing}), which does not have a direct analogue to radially-extended disks.  Although the ring theory $|\psi|$ oscillates in the bouncing regime (shown by the $\psi_0=0.2$ curve), it never decreases below $\psi_0$, whereas the simulated warps do.

\section{Discussion}
\label{sec:discussion}

\subsection{Turbulence}
\label{sec:parametric}

We have studied the onset of the bouncing regime and the accompanying nozzle shocks in warped disks.  However, disks usually exhibit magnetized turbulence. Even in the absence of magnetic fields, the parametric instability \citep{gammie_2000} of inviscid or nearly inviscid warps can generate turbulence even when the warp is small \citep{fairbairn_ogilvie_2023}, which damps bending waves \citep{deng_2021}. Regardless of the source of turbulence, it will result in an effective viscosity that damps shearing motions, such as sloshing. Since sloshing and warping multiply to drive breathing, any damping of the sloshing motions will increase the critical warp amplitude required for bouncing. In linear bending wave theory, the amplitude of the sloshing motions are usually determined by whichever is larger -- $\alpha$, $h$, or $\delta$ \citep{papaloizou_lin_1995}, where $\alpha$ is the effective viscosity parameter \citep{ss_73}. Since $\delta$ is $>0.1$ throughout the inner regions of a BH disk, only strong turbulent viscosities should significantly affect the bouncing criterion derived in this work. Our expectation that bouncing should persist in turbulent systems is bolstered by K23, where bouncing persisted in a global accretion disk subject to magnetized turbulence. 

\subsection{Broader implications}

A disk subject to bouncing and nozzle shocks behaves very differently from a planar disk. This was the case in K23, where nozzle shocks were the primary driver of accretion throughout most of the disk, dominating over magnetized turbulence. Warps are usually associated with BH disks when the disk has some tilt angle with respect to the BH spin axis. The tilt angle allows BH frame-dragging to force fluid parcels to precess differentially, resulting in a warp. If this warp is driven strongly enough, the disk will enter the bouncing regime. However, then nozzle shocks will rapidly dissipate the warp (e.g., Fig.~\ref{fig:warpevo}). 
Although our warps dissipated on several orbital timescales, nothing was driving them, so damping necessarily lead to their decay.
In reality, it is possible that disks are held at warp amplitudes very near the critical warp amplitude, and that nozzle shock dissipation and BH frame-dragging balance each other out. 


The energy dissipated per unit mass per orbit in a thin, viscous disk is $\approx h^2\alpha$, where $\alpha$ is the viscosity parameter \citep{pringle_1981}.
Using Equation \ref{eq:ediss_est}, we can then associate $|\eta|$ with an ``effective'' $\alpha$ viscosity parameter,
\begin{equation}
    \alpha \approx |\eta|^2
\end{equation}
Since $|\eta|>1$ in the bouncing regime, $\alpha>1$. This suggests that if accretion is driven by nozzle shocks, variability can occur on shorter timescales than a traditional $\alpha$ disk \citep{ss_73}, wherein $\alpha$ is both theoretically bounded below unity and observationally estimated to be $\approx0.1-0.4$ \citep[][]{king_2007}. K23 found effective $\alpha$ parameters $\gtrsim2-4$ in the outer regions of a strongly warped disk. We can use this to infer $|\eta|\approx2$, which leads to radial velocities,
\begin{equation}
v_r \approx -2\times10^{-3}\left(\frac{h}{0.02}\right)^2\left(\frac{|\eta|}{2}\right)^2v_{\rm K},
\label{eq:vrvk_nozzle}
\end{equation}
where $h=0.02$ is also consistent with K23. 
We can estimate the inflow timescale for the inner regions of a strongly warped supermassive black hole accretion disk by taking $t_{\rm inflow}\approx r/v_r$, finding 
\begin{equation}
    \begin{aligned}
    t_{\rm inflow} \approx 320\,\,&{\rm days}\left(\frac{M_{\rm SMBH}}{10^8\,M_\odot}\right)\left(\frac{r}{20\,r_{\rm g}}\right)^{3/2}\\\times&\left(\frac{h}{0.02}\right)^{-2}\left(\frac{|\eta|}{2}\right)^{-2}
    \end{aligned}
\label{eq:nozzle_inflow_time}
\end{equation}
For a quasar, this is a quite rapid inflow timescale, and is similar to the timescales associated with the extreme variability in changing-look AGN \citep{lawrence_2018,graham_2020,ricci_2023}. This may be a hint that some changing-look mechanisms are related to strongly warped accretion. 

 Nozzle shocks may be collisionless, as was recently considered in a similar scenario by \cite{sironi_2024}, who used particle-in-cell simulations to show that collisionless shocks appropriate to tilted thick disks can result in separate ion and electron temperatures. Since there are two nozzle shocks per orbit and the orbital timescale is very short, strong warps may result in two-temperature flows even in disks that are otherwise collisional (for instance, the soft state of X-ray binaries). Such a two-temperature state could have important dynamical and radiative consequences.

\subsection{Summary}
\label{sec:discussion:summary}
We have used both the analytic theory of warped rings and general-relativistic, hydrodynamic simulations of warped, radially-narrow tori to glean insights into nonlinear warps. Specifically, we have studied the onset of resonant bouncing above a critical warp amplitude and the nozzle shocks that result from it. Our main findings are,
\begin{itemize}
    \item Three-dimensional, general-relativistic simulations of radially-narrow, warped tori can be accurately modeled by a relativistic extension of the ring theory introduced by \cite{fairbairn_ogilvie_2021a}. We confirmed this by comparing our simulations with the linear modes of the ring theory, depicted in Figure \ref{fig:gentle_ring}. The three-dimensional simulations are able to capture these linear modes with excellent agreement, despite being initialized with different assumptions and within a global potential. This is true until the warp becomes sufficiently large.
    \item Tori with large enough warp amplitudes enter the ``bouncing regime'', characterized by large scale height oscillations twice an orbit. We derived the critical warp amplitude above which bouncing occurs, which in the non-Keplerian limit is $\psi_{\rm c}\approx0.4(|1-\kappa^2/\nu^2|)^{1/2}\approx(r/r_{\rm g})^{-1/2}$ and in the Keplerian limit is $\psi_{\rm c}\approx0.4\epsilon^2\approx0.4h^{1/2}$ (see Eq.~\ref{eq:psi_crit} for a more general expression). 
    This is in good agreement with our simulation results. While the ring theory predicts that these scale height oscillations can grow to very large values, we found in our simulations that the scale height oscillations are usually limited to amplitudes that are of order the equilibrium scale height. The atmosphere of the disk during a bounce can be vertically extended, with low density gas launched to heights that are 1-2 orders of magnitude larger than the equilibrium scale height.
    \item When disks are in the bouncing regime, they are subject to ``nozzle shocks'' twice an orbit. These nozzle shocks occur at the minima of the scale height oscillations. Each nozzle shock, as previously identified by K23, dissipates a significant fraction of the orbital energy of the torus in the simulated parameter space. We provide an estimate for the fractional dissipation in Equation \ref{eq:ediss_est}. We found that this dissipation can damp the warp to values below the critical warp amplitude within 10-20 orbits (Fig.~\ref{fig:warpevo}). We expect that nozzle shocks are able to drive inflow velocities that are much greater than standard $\alpha$ disks (Eq.~\ref{eq:vrvk_nozzle}). We also expect that in systems which have driven warps, rather than the freely-evolving warps studied in this work, nozzle shocks may regulate the bouncing regime such that the disk cannot greatly exceed the critical warp amplitude. 
\end{itemize}

While the ring theory describes the dynamics of radially-narrow tori, it can be related to radially-extended disks by taking the $\Delta r\rightarrow r$ limit, where $\Delta r$ is the ring half-width. In this limit, $\epsilon$ becomes $h$. The main caveat is that in a disk, the warp may also be torqued by neighboring annuli. However, it is generally expected that these interactions occur on inflow or sound crossing timescales, which are much longer than the orbital timescale on which the ring oscillations set in \citep[e.g., ][]{ogilvie_2022}. Although we have focused on accretion disks around black holes, where the general-relativistic potential drives the non-Keplerian behavior of the disk, our results are equally applicable to other types of disks as well. This is because the only role of the general-relativistic potential is to detune the epicyclic frequencies from the orbital frequency, and any manifestly relativistic effects are ordered out of the locally-expanded potential in the derivation of the ring theory in a Kerr metric (Appendix \ref{app:ring_theory}). Thus, our results may be applied to other types of non-Keplerian potentials -- such as from oblate stars or binary systems -- by making appropriate choices for the epicyclic frequencies $\kappa$ and $\nu$. Our results may also be applied to any Keplerian potential by taking the limit $\delta\rightarrow0$. 

In the future, more dedicated work needs to be done to ascertain which disks enter the bouncing regime, which likely depends most strongly on the BH spin, disk tilt, and disk aspect ratio. It is also necessary to test the warp amplitudes at which such disks saturate as this likely determines the accretion rate. If disks are so strongly warped that they tear, they likely pass through the bouncing regime first, so the question of disk tearing and nozzle shocks are inter-related. Finally, much of the physics outlined in this work should be tested in a more first-principles context, where magnetic fields, radiative cooling and collisionless shocks may all play a role depending on the astrophysical context that is envisioned.



\begin{acknowledgments}
We thank Callum Fairbairn, Gordon Ogilvie, Jonatan Jacquemin, Jiaru Li and Lorenzo Sironi for enlightening discussions. NK is supported by an NSF Graduate Research Fellowship. ML was supported by the John Harvard, ITC and NASA Hubble Fellowship Program fellowships, and NASA ATP award 21-ATP21-0077. An award of computer time was provided by the Innovative and Novel Computational Impact on Theory and Experiment (INCITE) and ASCR Leadership Computing Challenge (ALCC) programs under awards PHY129 and AST178. This research used resources of the Oak Ridge Leadership Computing Facility, which is a DOE Office of Science User Facility supported under Contract DE-AC05-00OR22725. Y.L. acknowledges NASA grant 80NSSC23K1262. AT acknowledges support by NASA 
80NSSC22K0031, 
80NSSC22K0799, 
80NSSC18K0565 
and 80NSSC21K1746 
grants, and by the NSF grants 
AST-2009884, 
AST-2107839, 
AST-1815304, 
AST-1911080, 
AST-2206471, 
AST-2407475, 
OAC-2031997. 
This research was supported in part by grant NSF PHY-2309135 to the Kavli Institute for Theoretical Physics (KITP).
\end{acknowledgments}

\appendix
\section{General-relativistic hydrostatic torus solution}
\label{app:gr_equilibrium}

Here, we  derive general-relativistic hydrodynamic equilibrium solutions for thin disks. These solutions can also model radially-narrow ``rings''. We use the relativistic von Zeipel theorem \citep{abramowicz_1971} to derive exact solutions to the general-relativistic Euler equation. This is a standard approach and we follow the steps outlined in \cite{chakrabarti_1985} and \cite{devilliers_2003}, except with a different angular momentum distribution. We use Boyer-Lindquist coordinates and the metric convention $(-\,+\,+\,+)$. We assume that only the azimuthal component of velocity is non-zero, e.g. the contravariant four-velocity reads,
\begin{equation}
u^\mu = (u^t,0,0,u^\varphi)
\label{eq:fourvelocity}
\end{equation}
The relativistic von Zeipel theorem states that, for barotropic equations of state $p=p(e)$ (where $p$ is the gas pressure and $e$ is the energy density), surfaces of constant specific angular momentum,
\begin{equation}
    l = -u_\varphi/u_t = -\frac{g_{t\varphi} + \Omega g_{\varphi\varphi}}{g_{tt} + \Omega g_{t\varphi}},
    \label{eq:l}
\end{equation}
and orbital frequency,
\begin{equation}
    \Omega = u^\varphi/u^t = -\frac{g_{t\varphi} + lg_{tt}}{g_{\varphi\varphi}+ lg_{t\varphi}},
    \label{eq:omega}
\end{equation}
coincide. These surfaces have the geometry of curved cylinders. We denote the "radii" of these curved cylinders as von Zeipel radii which we write as,
\begin{equation}
    \lambda^2 = l/\Omega = -\frac{lg_{\varphi\varphi}+l^2g_{t\varphi}}{g_{t\varphi} + lg_{tt}}
    \label{eq:lambda}
\end{equation}
Far from the central black hole, $\lambda$ asymptotically approaches the cylindrical radius. This relation, along with some prescribed von Zeipel relationship $l=l(\lambda)$, will allow us to determine our angular momentum distribution. We also assume that the torus is isentropic. We will write the spatial components of the relativistic Euler equation in the form given by Equation 9 of \cite{devilliers_2003}, 
\begin{equation}
    \frac{\partial_j h}{h} = -\frac{1}{2}\frac{\partial_j u_t^{-2}}{u_t^{-2}} + \frac{\Omega}{1 - l\Omega}\partial_j l
    \label{eq:euler}
\end{equation}
Here, $h=(p+e)/\rho$ is the specific enthalpy and $\rho$ is the gas density. We can integrate Eq.~\ref{eq:euler} provided a function $F(l)$  where $F^{-1}\partial_j F = \Omega/(1-l\Omega)\partial_j l$. This function $F(l)$ must satisfy the expression,
\begin{equation}
    {\rm ln\,F} = \int_{l(r_{\rm in})}^{l}-\frac{\Omega}{1-l\Omega} dl
    \label{eq:f_integrand}
\end{equation}
Here, $r_{\rm in}$ is the inner radius of the distribution and $l(r_{\rm in})$ is evaluated at the midplane ($\theta=\pi/2$). \cite{chakrabarti_1985} assumed $l$ obeys a power law with $\lambda$, which allowed them to analytically solve Eq.~\ref{eq:f_integrand}. However, we want to construct solutions which are very nearly Keplerian. Since the Keplerian angular momentum distribution does not obey a power law with $\lambda$, power-law formulations of $l(\lambda$) can only approximate Keplerian angular momentum profiles if the radial extent of the disk is $\Delta r \ll r_0$ where $\Delta r\equiv r_0-r_{\rm in}$. This is too restrictive, so we will instead assume the following von Zeipel relationship,
\begin{equation}
    l = l_{\rm c}(\lambda)\left[1 - \epsilon_{\rm torus}^2\left(\frac{\lambda-\lambda_0}{\lambda_0}\right)\right]
    \label{app:eq:amom_dist}
\end{equation}
Here, $\lambda_0$ is the von Zeipel radius corresponding to $r=r_0$ and $\theta=\pi/2$, where $r_0$ is the pressure maximum of the distribution. The parameter $\epsilon_{\rm torus}$ sets the aspect ratio of the disk. When $\Delta r < r$ (e.g., the disk is "radially-narrow"), $\epsilon_{\rm torus}$ is $\sim H/\Delta r$ and is analogous to the analytic aspect ratio $\epsilon$ of the ring structures derived in Appendix \ref{app:ring_theory}.
This distribution is exactly Keplerian at $\lambda_0$, sub-Keplerian at $\lambda>\lambda_0$, and super-Keplerian at $\lambda<\lambda_0$. We denote the specific angular momentum of circular orbits as $l_{\rm c}$, which satisfies the equation,
\begin{equation}
\partial_rg^{tt} - 2l_{\rm c}\partial_rg^{t\varphi} + l_{\rm c}^2\partial_rg^{\varphi\varphi} = 0
\label{eq:keplerian_angular_momentum}
\end{equation}
In general, Eq.~\ref{eq:keplerian_angular_momentum} has to be solved numerically. The quantity $l_{\rm c}(\lambda)$ is equal to the value of $l_{\rm c}$ that is on the midplane, $\theta = \pi/2$, of the von Zeipel cylinder $\lambda$. Since $l_{\rm c}$ is a function of $r$ and $r$ cannot be analytically solved in terms of $\lambda$, $l_{\rm c}$ must also be determined numerically. Given $l(\lambda)$, we can also determine $u_t(r,\theta)$ via Equation \ref{eq:fourvelocity} and the normalization condition $u^\mu u_\mu = -1$. With $F$ and $u_t$ determined everywhere, we can integrate Equation \ref{eq:euler} to find,
\begin{equation}
    h = \frac{u_t F}{u_t(r_{\rm in})f(r_{\rm in})}
    \label{eq:enthalpy}
\end{equation}
This function is valid within the potential surface defined at $r_{\rm in}$; outside of this surface, there is no gas. Our equilibrium solution is then completed by a choice of barotropic equation of state. We assume it to be polytropic, such that
\begin{equation}
e = np+\rho,
\label{eq:eos_e}
\end{equation}
where $n$ is the polytropic index, and
\begin{equation}
p = K\rho^{\gamma},
\label{eq:eos_p}
\end{equation}
where $K$ is a constant and $\gamma = 1+1/n$ is the adiabatic index. To recap, other than $\gamma$ our parameter choices are $r_{\rm in}$, which is the inner radius of the distribution; $r_0$, which is the radius of the pressure maximum; and $\epsilon_{\rm torus}$, which sets the aspect ratio of the disk. 

\section{Ring Theory in the Kerr metric}
\label{app:ring_theory}

Here, we rederive the Newtonian ring theory introduced in FO21a in the Kerr metric. The authors exactly solve the ideal, compressible fluid equations for fluid tori in a shearing sheet \citep{hawley_1995}. These tori are radially-narrow and we call them ``rings''. FO21a studied the oscillations of these rings and demonstrated a close correspondence with the dynamics of a warped accretion disk. However, their model is Newtonian, and we want to study rings embedded within the curved spacetime surrounding a black hole. We will do this by employing the general-relativistic shearing box equations derived in \cite{gammie_2004}.  

The Newtonian shearing box is described with local Cartesian coordinates $x=r-r_0$, $y=r_0(\phi-\Omega t)$ and $z=r{\rm sin}\theta$, where $\Omega$ is the orbital frequency of a circular reference orbit which is centered at $r=r_0$. In a Kerr metric, we can derive an analogous set of local, Cartesian coordinates by projecting our four-velocities onto the orthonormal tetrad carried by an observer on a circular reference orbit. This is done in \cite{novikov_thorne_1973} and we largely preserve their notation, except that we label the line metric in the orbiting frame with local Cartesian coordinates ($\tilde{x}$, $\tilde{y}$, $\tilde{z}$) and coordinate time $\tilde{\tau}$. The transformation from Boyer-Lindquist coordinates to the orbiting orthonormal tetrad $e^{\tilde{\nu}}_{\mu}$ carried is then,

\begin{align}
& e_\mu^{\tilde{t}} = (\frac{\mathcal{G}}{\mathcal{C}^{1/2}},0,0,-\frac{\mathcal{F}}{\mathcal{C}^{1/2}}r_0^{1/2})\\
& e_\mu^{\tilde{x}} = (0,\mathcal{D}^{-1/2},0,0)\\
& e_\mu^{\tilde{z}} = (0,0,-r_0,0)\\
& e_\mu^{\tilde{y}} = (-\frac{\mathcal{D}^{1/2}}{\mathcal{C}^{1/2}}r_0^{-1/2},0,0,\frac{\mathcal{B}\mathcal{D}^{1/2}}{\mathcal{C}^{1/2}}r_0)
\label{eq:tetrad}
\end{align}
Here, we have used the relativistic correction factors $\mathcal{B}=1 + a/r_0^{3/2}$, $\mathcal{C}=1 - 3/r_0 + 2a/r_0^{3/2}$, $\mathcal{D}=1 - 2/r_0 + a^2/r_0^2$, $\mathcal{F}=1 - 2a/r_0^{3/2} + a^2/r_0^2$ and $\mathcal{G}=1 - 2/r_0 + a/r_0^{3/2}$ which all approach unity as $r_0\rightarrow\infty$. The 1-forms in the orbiting basis are $dx^{\hat{\mu}}=e^{\hat{\mu}}_\nu dx^{\nu}$ and can be integrated to acquire a set of local coordinates, 
\begin{align}
    \label{eq:lc_coord:t}& \tilde{\tau} = t\mathcal{G}\mathcal{C}^{-1/2} - (\phi-\phi_0)r_0^{1/2}\mathcal{F}\mathcal{C}^{-1/2}\\
    \label{eq:lc_coord:x}& \tilde{x} = (r-r_0)\mathcal{D}^{-1/2}\\
    \label{eq:lc_coord:y}&\tilde{y}  = r_0(\phi-\phi_0-\Omega_0t)\mathcal{B}\mathcal{D}^{1/2}\mathcal{C}^{-1/2}\\
    \label{eq:lc_coord:z}& \tilde{z} = r_0{\rm cos}(\theta),
\end{align}
Here, $\Omega_0=d\phi/dt=1/(r_0^{3/2}+a)$, is the frequency of a circular orbit in the Boyer-Lindquist coordinate frame. From here on, we will drop the tilde from the local coordinates of the shearing box in the Kerr metric, e.g., $(\tilde{\tau},\tilde{x},\tilde{y},\tilde{z})\rightarrow(\tau,x,y,z)$. At the precise position of the reference orbit, the metric is exactly flat. \cite{gammie_2004} expanded the metric about the reference orbit to find,
\begin{equation}
    ds^2 = (-1 + s^2x^2 - \nu^2 z^2)
    d\tau^2 + 4\Omega x 
    d\tau dy + dx^2 + dy^2 + dz^2,
    \label{eq:local_metric}
\end{equation}
where $\Omega=r_0^{-3/2}$ is frequency at which the frame rotates. In this frame, the $s^2x^2 -\nu^2z^2$ term is responsible for centrifugal forces. The $g_{\tau y}$ term results from the rotation of the frame and is responsible for the Coriolis force. We have also introduced the tidal parameter, $s$,
\begin{equation}
    s^2 = \frac{3}{r_0^3}\left(\frac{\mathcal{D}}{\mathcal{C}}\right)
    \label{eq:tidal_parameter}
\end{equation}
As $r_0\rightarrow\infty$, $s^2\rightarrow -2r_0d_r\Omega\Omega$, which is the tidal parameter in the Newtonian shearing box. This is related to the orbital shear, $S$ via the expression
\begin{equation}
    S^2 = \frac{3}{4}s^2 = \frac{9}{4r_0^3}\left(\frac{\mathcal{D}}{\mathcal{C}}\right),
    \label{eq:app:orbital_shear}
\end{equation}
which moving forward we will use instead of $s$. We have also introduced the vertical epicyclic frequency, $\nu$, which measured in the orbiting frame is,
\begin{equation}
    \nu^2 = \frac{1}{r_0^3}\left(\frac{1-4ar_0^{-3/2}+3a^2r_0^{-2}}{\mathcal{C}}\right)
    \label{eq:app:ff_vertical_epicyclic}
\end{equation}
For later reference, we also write down the radial epicyclic frequency as measured in the orbiting frame,
\begin{equation}
    \kappa^2 = \frac{1}{r_0^3}\left(\frac{1 - 6r_0^{-1} + 8ar_0^{-3/2} - 3a^2r_0^{-2}}{\mathcal{C}}\right)
    \label{eq:app:ff_radial_epicyclic}
\end{equation}
Now, we make the assumption that within the orbiting frame the fluid velocities are non-relativistic, with the following justification. The oscillation frequencies of a ring can reach $\mathcal{O}(\Omega)$ when the warp is large. If we regard the width of a ring as $L$, then the maximum velocities within the ring are $\mathcal{O}(\Omega L)$. So, as long as $L\ll r_0$, the fluid velocities in the orbiting frame will be approximately non-relativistic even very close to the black hole where $r_0\Omega$ approaches $c$. In this approximation, we can neglect the difference between contravariant and covariant indices, and so we will stick to Newtonian notation for the remainder of this section. The non-relativistic equations of motion in the orbiting frame are then,
\begin{align}
\label{eq:shearing_box_eom1}
&D_{\tau}v_x - 2\Omega v_y = 4S^2x/3 - \frac{1}{\rho}\partial_{x}p\\
\label{eq:shearing_box_eom2}
&D_{\tau}v_y + 2\Omega v_x = - \frac{1}{\rho}\partial_{y}p\\
\label{eq:shearing_box_eom3}
&D_{\tau}v_z = -\nu^2z - \frac{1}{\rho}\partial_{z}p,
\end{align}
where $D_\tau=\partial_\tau + v_x\partial_x + v_y\partial_y + v_z\partial_z$. These are essentially the same as the equations of motion in a Newtonian shearing box and can be compared to Eqs. 1-3 of FO21a. The only distinctions are that the shear, $S$, and the vertical epicyclic frequency, $\nu$, include relativistic correction factors. We can now proceed with a summary of the FO21a ring model except with these slight modifications.

The simplest solution of Equations \ref{eq:shearing_box_eom1}-\ref{eq:shearing_box_eom3} corresponds to a shear flow of circular orbits. This is defined by an azimuthal velocity profile $\vec{v}=-2S^2/(3\Omega)x\hat{e}_y$. This velocity profile results in a flow structure that is infinite in radial extent and has no radial pressure gradients. We instead assume the velocity profile $\vec{v} = - Ax\hat{e}_y$ where $A>2S^2/(3\Omega)$, which establishes a geostrophic balance between the Coriolis force and nonzero radial pressure gradients, with sub-"Keplerian" flow at radii $>r_0$ and super-"Keplerian" flow at radii $<r_0$. The resulting equilibrium structure consists of density and pressure contours that are elliptical in the coordinates $x$ and $z$. The exact value of $A$ sets the aspect ratio, $\epsilon$, of the ellipse,
\begin{equation}
    \epsilon = \sqrt{\frac{2\Omega_0A-4S^2/3}{\nu^2}},
    \label{eq:fa_epsilon}
\end{equation}


The elliptical, equilibrium ring can be generalized to dynamically oscillating rings by assuming a flow field that is linear in local Cartesian coordinates, 
\begin{equation}
    v_i = A_{ij}x_{j}
    \label{eq:fairbairn_flow_matrix}
\end{equation}
where $x_j \equiv (x,\,y,\,z)$ and Latin indices span $i=1,2,3$. This flow field is independent of azimuth, i.e. $A_{i2} = 0$. Here, $A_{ij}$ is the "flow matrix". The equilibrium solution occurs when the only non-zero component of the flow matrix is $A_{21}>2S^2/(3\Omega)$. By allowing the other components of $A_{ij}$ to be non-zero, a variety of dynamical behavior can be captured, such as breathing motions (e.g., $A_{11}\neq0$, $A_{33}\neq0$) and tilting motions (e.g., $A_{13}\neq0$, $A_{31}\neq0$). FO21a also assumed a materially invariant function $f(x,z,t)$ and that the density and pressure are separable into the terms $\rho=\hat{\rho}(t)\tilde{\rho}(f)$ and $p=\hat{p}(t)\tilde{p}(f)$. They adopt the following functional form of $f$, 
\begin{equation}
    f = C - \frac{1}{2}S_{ij}x_ix_j
    \label{eq:app:shape_matrix}
\end{equation}
Here, $C$ is a constant and $S_{ij}(t)$ is a time dependent, positive-definite "shape matrix" where $S_{i2}=S_{2i}=0$. We can see that a diagonal $S_{ij}$ results in an elliptical torus with reflection symmetry about the midplane, while off-diagonal components result in corrugations of the torus that break this symmetry. The shape matrix is related to the aspect ratio of their ellipse via the relation $\epsilon=\sqrt{S_{11}/S_{33}}$. Finally, the authors also define the characteristic temperature $\hat{T} = \hat{p}/\hat{\rho}$. Using these definitions, they derive from the ideal, compressible fluid equations the following set of ten first order, coupled, non-linear ordinary differential equations,
\begin{align}
    \label{eq:fa:1}&d_tS_{11} + 2(S_{11}A_{11}+S_{13}A_{31})=0\\
    \label{eq:fa:2}&d_tS_{13} + S_{11}A_{13} + S_{33}A_{31} + S_{13}(A_{11}+A_{33})=0\\
    \label{eq:fa:3}&d_tS_{33} + 2(S_{13}A_{13} + S_{33}A_{33})=0\\
    \label{eq:fa:4}&d_tA_{11} + A_{11}^2 + A_{13}A_{31} - 2\Omega A_{21} =4S^2/3 + \hat{T}S_{11}\\
    \label{eq:fa:5}&d_tA_{13} + A_{11}A_{13} + A_{13}A_{33} - 2\Omega A_{23} = \hat{T}S_{13}\\
    \label{eq:fa:6}&d_tA_{21} + A_{21}A_{11} + A_{23}A_{31} + 2\Omega A_{11}=0\\
    \label{eq:fa:7}&d_tA_{23} + A_{21}A_{13} + A_{23}A_{33} + 2\Omega A_{13} = 0\\
    \label{eq:fa:8}&d_tA_{31} + A_{31}A_{11} + A_{33}A_{31} = \hat{T}S_{13}\\
    \label{eq:fa:9}&d_tA_{33} + A_{31}A_{13} + A_{33}^2 = -\nu^2 + \hat{T}S_{33}\\
    \label{eq:fa:10}&d_t\hat{T}=-(\gamma-1)\hat{T}(A_{11}+A_{33})
\end{align}
The most relevant terms for the purpose of this work are $A_{31}\equiv\psi\nu$, $A_{13}\equiv\sigma\nu$ and $A_{33}\equiv\eta\nu$, and we use the dimensionless variables $\psi$, $\sigma$ and $\eta$ throughout the main text. These equations are identical to those derived in FO21a except for the relativistic corrections to the vertical epicyclic frequency, $\nu$, and the orbital shear, $S$. Approximate forms of these relativistic frequencies are provided in Equation \ref{eq:apprx_relativistic_freqs}. Equations \ref{eq:fa:1}-\ref{eq:fa:3} describe the evolution of the shape of the torus, Equations \ref{eq:fa:4}-\ref{eq:fa:9} describe the evolution of the flow field, and Equation \ref{eq:fa:10} describes the evolution of the temperature. In the equilibrium state, the only non-zero component of the flow matrix is $A_{21}$,
\begin{equation}
    A_{21}^{\rm (eq)} = -\frac{1}{2\Omega}\left(4S^2/3+\epsilon^2\nu^2\right)
\end{equation}
The only non-zero components of the shape matrix are $S_{33}$ and $S_{11}$, which describe the inverse squaraes of the semiminor and semimajor axes of the ring, respectively,
\begin{equation}
\begin{aligned}
    &S_{33}^{\rm (eq)} = \nu^{-2}\hat{T}^{\rm (eq)} \\
    &S_{11}^{\rm (eq)} = \epsilon^2S_{33}^{(\rm eq)}
\end{aligned}
\end{equation}
where the equilibrium temperature is
\begin{equation}
    \hat{T}^{\rm (eq)} = \epsilon^2\nu^2
\end{equation}


\section{Linear modes of ring equations}
\label{app:linear_modes}
Here, we write down the linear modes of the ring equations. We can split the ring equations (Eqs. \ref{eq:fa:1}-\ref{eq:fa:10}) into two classes of modes, which decouple at lowest order: ``tilting'' modes and ``breathing'' modes. This is the same linear analysis as was done in FO21a, except with relativistic corrections due to the Kerr metric.
\subsection{Tilting Modes}
\label{app:linear_modes:warping}
We can calculate tilting modes by introducing small perturbations to the quantities which break the midplane symmetry of the ring ($A_{31}$, $A_{13}$, $A_{23}$ and $S_{13}$ in Eqs. \ref{eq:fa:1}-\ref{eq:fa:10}) and assuming a time-dependence ${\rm exp}(i\omega_\psi t)$ for each term. We denote each perturbed quantity with a prime (e.g., $X'$) and each unperturbed quantity without (e.g., $X$). The perturbed equations read, 
\begin{equation}
\begin{aligned}
    &i\omega_\psi S_{13}' + S_{11}A_{13}' + S_{33}A_{31}'=0 \\
    &i\omega_\psi A_{13}' - 2\Omega A_{23}' = \hat{T}S_{13}' \\
    &i\omega_\psi A_{23}' + A_{21}A_{13}' + 2\Omega A_{13}' = 0 \\
    &i\omega_\psi A_{31}' = \hat{T}S_{13}' 
\end{aligned}
\end{equation}
These can be further reduced to two equations for $A_{13}'$ and $A_{31}'$,
\begin{equation}
\begin{aligned}
    \omega_\psi^2&A_{13}' = \kappa^2 A_{13}' + \nu^2 A_{31}'\\
    \omega_\psi^2&A_{31}' = \nu^2(\epsilon^2A_{13}' + A_{31}'),
\end{aligned}
\end{equation}
where we have used the relations $S_{11}/S_{33}=\epsilon^2$ and $\kappa^2 = 2\Omega A_{21} + 4\Omega^2 + \epsilon^2\nu^2$. The resulting linear eigenvalue problem is,
\begin{equation}
    \omega_\psi^2
    \begin{pmatrix}
    A_{13}'\\A_{31}'
    \end{pmatrix} =
    \begin{pmatrix}
    \kappa^2&\nu^2\\
    \nu^2\epsilon^2&\nu^2\\
    \end{pmatrix}
    \begin{pmatrix}
    A_{13}'\\A_{31}'
    \end{pmatrix}
    \label{app:eq:linear_warping_eigenproblem},
\end{equation}
The resulting eigenvalues are, 
\begin{equation}
    \omega_{\psi,\pm}^2 = \frac{1}{2}    ~\left(\kappa^2 + \nu^2\pm\sqrt{(\kappa^2-\nu^2)^2 + 4\epsilon^2\nu^4}\right).
    \label{eq:app:warping_eigenvalues}
\end{equation}
This is the same eigenvalue relation derived in the Newtonian
case by FO21a. The associated eigenvector relationship is,
\begin{equation}
    \begin{aligned}
    \begin{pmatrix}
    A_{13}'\\A_{31}'
    \end{pmatrix}
    =\,
    &c_-
    \begin{pmatrix}
    2\left(\delta-\sqrt{\delta^2 + 4\epsilon^2}\right)^{-1}\\1
    \end{pmatrix}{\rm exp}(i\omega_{\psi,-}t)
    \\+\,&c_+
    \begin{pmatrix}
    2\left(\delta+\sqrt{\delta^2 + 4\epsilon^2}\right)^{-1}\\1
    \end{pmatrix}{\rm exp}(i\omega_{\psi,+}t),
    \end{aligned}
    \label{eq:app:warping_eigenvectors}
\end{equation}
These linear tilting modes are the ring analogues of bending waves in global accretion disks.

\subsection{Breathing Modes}
\label{app:linear_modes:breathing}
We calculate breathing modes by introducing small perturbations to the quantities which preserve the midplane symmetry of the ring ($A_{11}$, $A_{33}$, $A_{21}$, $S_{11}$, $S_{33}$, and $\hat{T}$ in Eqs. \ref{eq:fa:1}-\ref{eq:fa:10}) and assuming a time-dependence ${\rm exp}(i\omega_\eta t)$ for each term. We will denote each perturbed quantity with a prime (e.g., $X'$) and each unperturbed quantity without (e.g., $X$). The perturbed equations read, 
\begin{equation}
\begin{aligned}
&i\omega_\eta S_{11}' + 2S_{11}A_{11}' = 0 \\
&i\omega_\eta S_{33}' + 2S_{33}A_{33}' = 0 \\
&i\omega_\eta A_{11}' - 2\Omega A_{21}' = \hat{T}'S_{11} + \hat{T}S_{11}' \\
&i\omega_\eta A_{21}' + A_{21}A_{11}' + 2\Omega A_{11}' = 0 \\
&i\omega_\eta A_{33}' = \hat{T}'S_{33} + \hat{T}S_{33}' \\
&i\omega_\eta \hat{T}' = -(\gamma - 1)\hat{T}(A_{11}' + A_{33}')
\end{aligned}
\end{equation}
This can be further reduced to two equations for $A_{11}'$ and $A_{33}'$,
\begin{equation}
\begin{aligned}
&\omega_\eta^2 A_{11}' = (\kappa^2 + \gamma\epsilon^2\nu^2)A_{11}' + \epsilon^2\nu^2(\gamma-1)A_{33}'\\
&\omega_\eta^2 A_{33}' = \nu^2(\gamma-1)A_{11}' + \nu^2(\gamma+1)A_{33}',
\end{aligned}
\label{eq:app:linear:breathing:linearized}
\end{equation}
where we have used the relations $S_{11}/S_{33}=\epsilon^2$ and $\kappa^2 = 2\Omega A_{21} + 4\Omega^2 + \epsilon^2\nu^2$. The resulting linear eigenvalue problem is,
\begin{equation}
    \omega_\eta^2
    \begin{pmatrix}
    A_{11}'\\A_{33}'
    \end{pmatrix} =
    \begin{pmatrix}
    \kappa^2+\gamma\epsilon^2\nu^2&\epsilon^2\nu^2(\gamma-1)\\
    \nu^2(\gamma-1)&\nu^2(\gamma+1)\\
    \end{pmatrix}
    \begin{pmatrix}
    A_{11}'\\A_{33}'
    \end{pmatrix}
    \label{eq:app:linear_breathing_eigenproblem},
\end{equation}
which can be solved to find the following eigenvalues,
\begin{equation}
    \omega_{\eta,\pm}^2 = \frac{1}{2}\left[\kappa^2 + (1 + \gamma + \gamma\epsilon^2)\nu^2 \pm \sqrt{(\kappa^2 + (1 + \gamma + \gamma\epsilon^2)\nu^2)^2 - 4((1+\gamma)\kappa^2\nu^2 + (3\gamma-1)\epsilon^2\nu^4)}\right],
    \label{eq:app:breathing_eigenvalues}
\end{equation}
which are the same as those derived in the Newtonian case by FO21a. This is an expression for four out of the six eigenmodes that exist for the sixth order system; the two remaining modes have zero frequency. The corresponding set of eigenvectors are,

\begin{equation}
\begin{aligned}
    \begin{pmatrix}
    A_{11}'\\A_{33}'
    \end{pmatrix}
    &=     c_1\begin{pmatrix}
    \frac{\kappa^2 + \nu^2 \left(\gamma\left(\epsilon^2-1\right)-1\right)+\sqrt{2 \kappa^2 \nu^2 \left(\gamma\left(\epsilon^2-1\right)
   -1\right)+\nu^4\left(\epsilon^4 \gamma^2+2 \epsilon^2((\gamma-5)\gamma+2)+(\gamma +1)^2\right)+\kappa^4}}{2
   (\gamma-1) \nu^2}\\1
    \end{pmatrix}{\rm exp}(i\omega_{\eta,+}t)\\
    &+ c_2\begin{pmatrix}
    \frac{\kappa^2 + \nu^2 \left(\gamma\left(\epsilon^2-1\right)-1\right)-\sqrt{2 \kappa^2 \nu^2 \left(\gamma\left(\epsilon^2-1\right)
   -1\right)+\nu^4\left(\epsilon^4 \gamma^2+2 \epsilon^2((\gamma-5)\gamma+2)+(\gamma +1)^2\right)+\kappa^4}}{2
   (\gamma-1) \nu^2}\\1
    \end{pmatrix}{\rm exp}(i\omega_{\eta,-}t)
\end{aligned}
    \label{eq:app:breathing_eigenvectors}
\end{equation}


\section{Quasi-linear forced breathing}
\label{app:ql_forced_breathing}

In this section, we will use the set of ring equations described in Appendix \ref{app:ring_theory} to derive the response of the breathing motions, $A_{33}$, to the forcing by linear tilting modes. This can be regarded as ``quasi-linear'' as we use the linear tilting modes (Appendix \ref{app:linear_modes:warping}) and input them as second-order forcing terms which modify the linear breathing modes (Appendix \ref{app:linear_modes:breathing}). 

The linear tilting modes as first order, which we indicate with a single prime (e.g., $X'$). These have time dependence ${\rm exp}(i\omega_\psi t)$. This set of perturbed quantities includes $S_{13}'$, $A_{13}'$, $A_{23}'$ and $A_{31}'$. We will write second-order perturbed quantities as $X''$. This set includes $S_{11}''$, $S_{33}''$, $A_{11}''$, $A_{33}''$, $A_{21}''$ and $\hat{T}''$. Since we are interested in second-order behavior, we will only keep linear tilting mode terms that result in nonlinear combinations of the form $X'Y'$. 

We first input these perturbed quantities into Equations \ref{eq:fa:1}-\ref{eq:fa:10}, 

\begin{align}
d_tS_{11}'' + 2S_{11}A_{11}''  &= -2S_{13}'A_{31}' \\
d_t S_{33}'' + 2S_{33}A_{33}'' &= -2S_{13}'A_{13}' \\
d_t A_{11}'' - 2\Omega A_{21}'' - \hat{T}''S_{11} - \hat{T}S_{11}''&= -A_{13}'A_{31}'\\
d_t A_{21}'' + A_{21}A_{11}'' + 2\Omega A_{11}''&=-A_{23}'A_{31}'\\
d_t A_{33}'' - \hat{T}''S_{33} - \hat{T}S_{33}'' &= -A_{31}'A_{13}' \\
d_t \hat{T}'' + (\gamma-1)\hat{T}(A_{11}'' + A_{33}'')&= 0
\end{align}
We will simplify this by substituting the linear order dependence on $S_{13}'$ and $A_{23}'$ with $A_{13}'$ and $A_{31}'$, using the relations
\begin{align}
    \label{eq:s13_linear}i\omega_\psi S_{13}' + S_{11}A_{13}' + S_{33}A_{31}' &=0\\
    \label{eq:a23_linear}i\omega_\psi A_{23}' + A_{21}A_{13}' + 2\Omega A_{13}'&=0
\end{align}
which results in the following expressions,
\begin{align}
d_tS_{11}'' + 2S_{11}A_{11}'' &= -\frac{2i}{\omega_\psi}\left(S_{11}A_{13}' + S_{33}A_{31}'\right)A_{31}' \\
d_t S_{33}'' + 2S_{33}A_{33}'' &= -\frac{2i}{\omega_\psi}\left(S_{11}A_{13}' + S_{33}A_{31}'\right)A_{13}' \\
d_t A_{11}'' - 2\Omega A_{21}'' - \hat{T}''S_{11} - \hat{T}S_{11}''&= -A_{13}'A_{31}'\\
d_t A_{21}'' + A_{21}A_{11}'' + 2\Omega A_{11}''&=-\frac{i}{\omega_\psi}\left(A_{21}A_{13}' + 2\Omega A_{13}'\right)A_{31}'\\
d_t A_{33}'' - \hat{T}''S_{33} - \hat{T}S_{33}''&= -A_{31}'A_{13}'\\
d_t \hat{T}'' + (\gamma-1)\hat{T}(A_{11}'' + A_{33}'')&= 0
\end{align}

This can be further simplified in the limit of small $\epsilon$. Here, we recall that in equilibrium, $S_{11}=\epsilon^2S_{33}$. So, when $\epsilon$ is small, all $S_{11}$ terms are ordered out. We also assume ``$u_z$-dominated'' linear tilting modes (Eq.~\ref{eq:linear_warping_eigenvalues}) such that $\omega_\psi\approx\nu$. Then, we arrive at a single equation for $A_{11}''$,
\begin{equation}
    d_{t}^2A_{11}'' + \kappa^2 A_{11}'' = -i\left(\frac{\kappa^2}{\nu} + 2\nu\right)A_{13}'A_{31}'
    \label{eq:app:forced_a11}
\end{equation}
which describes linear, forced oscillations of the radial breathing of the ring. Indeed, the natural frequency is the radial epicyclic frequency $\kappa$, as in the ``$u_x$-dominated'' breathing mode in Eq.~\ref{eq:linear_breathing_eigenvalues}. We can similarly solve for an equation for $A_{33}''$,
\begin{equation}
    d_t^2A_{33}'' + (\gamma+1)\nu^2 A_{33}'' + (\gamma-1)\nu^2A_{11}'' = -i4\nu A_{13}'A_{31}'
    \label{eq:app:forced_a33}
\end{equation}
Here, we must input a solution for $A_{11}''$. We will solve for the forced response of $A_{11}''$ by assuming a time dependence ${\rm exp}(i2\omega_\psi t)$ in Eq.~\ref{eq:app:forced_a11}, which yields
\begin{equation}
    A_{11}'' = \frac{i}{\nu}\left(\frac{3-\delta}{3+\delta}\right)A_{13}'A_{31}'\approx\frac{i}{\nu}(1-2\delta/3)A_{13}'A_{31}',
\end{equation}
where we have substituted $\kappa^2=\nu^2(1-\delta)$, where $\delta$ is the ``deviation from Keplerian resonance'' (Eq.~\ref{eq:deviation_resonance}) which we have assumed to be small. From this, we can see that $A_{11}''$ contributes to the forcing of $A_{33}''$ in Eq.~\ref{eq:app:forced_a33}, 
\begin{equation}
    d_t^2A_{33}'' + (\gamma+1)\nu^2A_{33}'' = -i\nu\left[(\gamma+3) - 2\delta/3(\gamma-1)\right]A_{13}'A_{31}'
\end{equation}
We will now return to the dimensionless notation used in the main text ($A_{13}'=\sigma\nu$, $A_{31}'=\psi\nu$, $A_{33}''=\eta\nu$) to rewrite Eq.~\ref{eq:app:forced_a11},
\begin{equation}
    d_t^2\eta + (\gamma+1)\nu^2\eta = -iF_\eta{\rm exp}(i2\omega_\psi t),
\end{equation}
such that the warp-induced driving force has amplitude,
\begin{equation}
    F_\eta \equiv \nu^2\left[(\gamma+3) - 2\delta/3(\gamma-1)\right]|\psi\sigma|,
    \label{eq:app:forcing_amplitude}
\end{equation}
and the general solution to $\eta$ is,
\begin{equation}
    \eta = \eta^{\rm (n)}{\rm exp}(i(\omega_\eta t + \phi)) + \frac{F_\eta}{\omega_\psi^2-\omega_\eta^2}{\rm exp}(i2\omega_\psi t),
    \label{eq:app:forced_breathing_solution}
\end{equation}
where $\phi$ is the phase and $\eta^{\rm (n)}$ is the amplitude of the ``natural'' (homogeneous) solution which oscillates at $\omega_\eta=\sqrt{\gamma+1}\nu$. We can also replace $\sigma$ in Eq.~\ref{eq:app:forcing_amplitude} by choosing an appropriate eigenvector that relates it to $\psi$. Per our discussion in Section \ref{sec:rings:linear:warp}, we regard the ``$u_z$-dominated'' mode (Eq.~\ref{eq:linear_warping_eigenvalues}) as the relevant mode. Taking the ``+'' eigenvector from Eq.~\ref{eq:app:warping_eigenvectors} (i.e., using $c_+=1$ and $c_-=0$), we derive our final expression for the forcing response of $\eta$,

\begin{figure}
    \centering
    \includegraphics[width=0.75\textwidth]{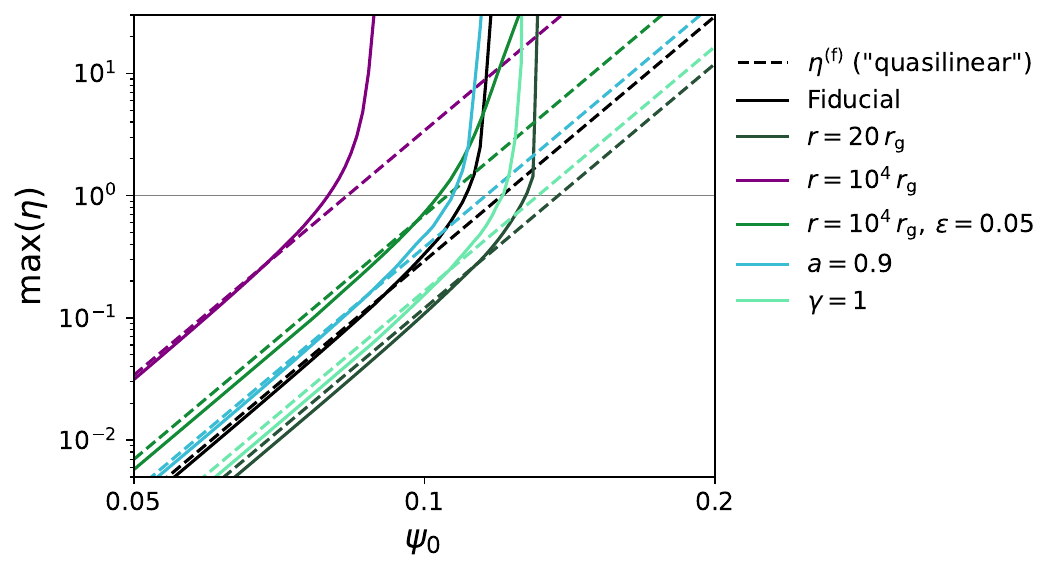}
\caption{Our quasi-linear estimate for the breathing amplitude (Eq.~\ref{eq:app:ql_breathing}) is accurate until the warp amplitude is large enough such that bouncing becomes imminent. As in Figure \ref{fig:bouncing1d}, we depict the maximum value of $\eta$ as a function of initial warp amplitude $\psi_0$ for integrations of the ring equations (solid lines) and our quasi-linear predictions (dashed lines). The "fiducial" (black) curve uses the same parameters as in the main body of the work - $r_0=50\,r_{\rm g}$, $a=0$, $\gamma=5/3$, $\epsilon=0.01$. Each of the other curves change one of these parameters, as labeled.}
\label{fig:bouncing1d_app}
\end{figure}

\begin{equation}
    \eta^{\rm (f)} \equiv \frac{F_\eta}{\omega_\psi^2-\omega_\eta^2}\approx |\psi|^2\frac{2\gamma+6 - 4\delta/3(\gamma-1)}{\bigl(\gamma-3\bigl)\bigl(\delta+\sqrt{\delta^2 + 4\epsilon^2}\bigl)}
    \label{eq:app:ql_breathing}
\end{equation}
In the Keplerian ($\delta \ll \epsilon$) limit, this expression scales as $\propto |\psi|^2/\epsilon$, which is consistent with FO21b, and in the non-Keplerian limit ($\epsilon \ll \delta$) this expression scales as $\propto |\psi|^2/\delta$. Both limits have an order-unity dependence on the adiabatic index.

We compare Equation \ref{eq:app:ql_breathing} to integrations of the ring equations in Figure \ref{fig:bouncing1d_app}, where we have plotted the maximum breathing amplitude over a course of an integration, ${\rm max}(\eta)$, as a function of initial warp amplitude $\psi_0$. This is analogous to Fig.~\ref{fig:bouncing1d} except for a wider parameter space. In all cases shown here, we have initialized the rings with vertical tilting modes (``+'' solutions in Eq.~\ref{eq:linear_warping_eigenvectors}). The ``fiducial'' integration (black curve) has the same parameters used throughout this work: $r_0=50\,r_{\rm g}$, $a=0$, $\epsilon=0.01$, $\gamma=5/3$, and in each of the other integrations we have changed one of these quantities. For each curve, our estimate for $\eta^{\rm (f)}$ (dashed lines) accurately predicts the behavior of the ring equations in the quasi-linear regime. As the breathing motions become nonlinear, the quasi-linear estimation is no longer good, which is expected. 

\bibliographystyle{aasjournal}
\bibliography{references}

\begin{thebibliography}{}
\expandafter\ifx\csname natexlab\endcsname\relax\def\natexlab#1{#1}\fi
\providecommand{\url}[1]{\href{#1}{#1}}
\providecommand{\dodoi}[1]{doi:~\href{http://doi.org/#1}{\nolinkurl{#1}}}
\providecommand{\doeprint}[1]{\href{http://ascl.net/#1}{\nolinkurl{http://ascl.net/#1}}}
\providecommand{\doarXiv}[1]{\href{https://arxiv.org/abs/#1}{\nolinkurl{https://arxiv.org/abs/#1}}}

\bibitem[{{Abramowicz}(1971)}]{abramowicz_1971}
{Abramowicz}, M.~A. 1971, \actaa, 21, 81

\bibitem[{{Bardeen} \& {Petterson}(1975)}]{bardeen_petterson_1975}
{Bardeen}, J.~M., \& {Petterson}, J.~A. 1975, \apjl, 195, L65, \dodoi{10.1086/181711}

\bibitem[{{Bollimpalli} {et~al.}(2024){Bollimpalli}, {Fragile}, {Dewberry}, \& {Klu{\'z}niak}}]{deepika_2024}
{Bollimpalli}, D.~A., {Fragile}, P.~C., {Dewberry}, J.~W., \& {Klu{\'z}niak}, W. 2024, \mnras, 528, 1142, \dodoi{10.1093/mnras/stad3975}

\bibitem[{{Bollimpalli} {et~al.}(2023){Bollimpalli}, {Fragile}, \& {Klu{\'z}niak}}]{deepika_2023}
{Bollimpalli}, D.~A., {Fragile}, P.~C., \& {Klu{\'z}niak}, W. 2023, \mnras, 520, L79, \dodoi{10.1093/mnrasl/slac155}

\bibitem[{{Chakrabarti}(1985)}]{chakrabarti_1985}
{Chakrabarti}, S.~K. 1985, \apj, 288, 1, \dodoi{10.1086/162755}

\bibitem[{{Chatterjee} {et~al.}(2019){Chatterjee}, {Liska}, {Tchekhovskoy}, \& {Markoff}}]{koushik_2019}
{Chatterjee}, K., {Liska}, M., {Tchekhovskoy}, A., \& {Markoff}, S.~B. 2019, \mnras, 490, 2200, \dodoi{10.1093/mnras/stz2626}

\bibitem[{{De Villiers} {et~al.}(2003){De Villiers}, {Hawley}, \& {Krolik}}]{devilliers_2003}
{De Villiers}, J.-P., {Hawley}, J.~F., \& {Krolik}, J.~H. 2003, \apj, 599, 1238, \dodoi{10.1086/379509}

\bibitem[{{Debes} {et~al.}(2017){Debes}, {Poteet}, {Jang-Condell}, {Gaspar}, {Hines}, {Kastner}, {Pueyo}, {Rapson}, {Roberge}, {Schneider}, \& {Weinberger}}]{debes_2017}
{Debes}, J.~H., {Poteet}, C.~A., {Jang-Condell}, H., {et~al.} 2017, \apj, 835, 205, \dodoi{10.3847/1538-4357/835/2/205}

\bibitem[{{Deng} \& {Ogilvie}(2022)}]{deng_ogilvie_2022}
{Deng}, H., \& {Ogilvie}, G.~I. 2022, \mnras, 512, 6078, \dodoi{10.1093/mnras/stac858}

\bibitem[{{Deng} {et~al.}(2021){Deng}, {Ogilvie}, \& {Mayer}}]{deng_2021}
{Deng}, H., {Ogilvie}, G.~I., \& {Mayer}, L. 2021, \mnras, 500, 4248, \dodoi{10.1093/mnras/staa3504}

\bibitem[{{Dullemond} {et~al.}(2022){Dullemond}, {Kimmig}, \& {Zanazzi}}]{dullemond_2022}
{Dullemond}, C.~P., {Kimmig}, C.~N., \& {Zanazzi}, J.~J. 2022, \mnras, 511, 2925, \dodoi{10.1093/mnras/stab2791}

\bibitem[{{Evans} \& {Kochanek}(1989)}]{evans_kochanek_1989}
{Evans}, C.~R., \& {Kochanek}, C.~S. 1989, \apjl, 346, L13, \dodoi{10.1086/185567}

\bibitem[{{Fairbairn} \& {Ogilvie}(2021{\natexlab{a}})}]{fairbairn_ogilvie_2021a}
{Fairbairn}, C.~W., \& {Ogilvie}, G.~I. 2021{\natexlab{a}}, \mnras, 505, 4906, \dodoi{10.1093/mnras/stab1554}

\bibitem[{{Fairbairn} \& {Ogilvie}(2021{\natexlab{b}})}]{fairbairn_ogilvie_2021b}
---. 2021{\natexlab{b}}, \mnras, 508, 2426, \dodoi{10.1093/mnras/stab2717}

\bibitem[{{Fairbairn} \& {Ogilvie}(2023)}]{fairbairn_ogilvie_2023}
---. 2023, \mnras, 520, 1022, \dodoi{10.1093/mnras/stad211}

\bibitem[{{Fragile} \& {Blaes}(2008)}]{fragile_blaes_2008}
{Fragile}, P.~C., \& {Blaes}, O.~M. 2008, \apj, 687, 757, \dodoi{10.1086/591936}

\bibitem[{{Gammie}(2004)}]{gammie_2004}
{Gammie}, C.~F. 2004, \apj, 614, 309, \dodoi{10.1086/423443}

\bibitem[{{Gammie} {et~al.}(2000){Gammie}, {Goodman}, \& {Ogilvie}}]{gammie_2000}
{Gammie}, C.~F., {Goodman}, J., \& {Ogilvie}, G.~I. 2000, \mnras, 318, 1005, \dodoi{10.1046/j.1365-8711.2000.03669.x}

\bibitem[{{Graham} {et~al.}(2020){Graham}, {Ross}, {Stern}, {Drake}, {McKernan}, {Ford}, {Djorgovski}, {Mahabal}, {Glikman}, {Larson}, \& {Christensen}}]{graham_2020}
{Graham}, M.~J., {Ross}, N.~P., {Stern}, D., {et~al.} 2020, \mnras, 491, 4925, \dodoi{10.1093/mnras/stz3244}

\bibitem[{{Greenhill} {et~al.}(1995){Greenhill}, {Henkel}, {Becker}, {Wilson}, \& {Wouterloot}}]{greenhill_1995}
{Greenhill}, L.~J., {Henkel}, C., {Becker}, R., {Wilson}, T.~L., \& {Wouterloot}, J.~G.~A. 1995, \aap, 304, 21

\bibitem[{{Greenhill} {et~al.}(2003){Greenhill}, {Booth}, {Ellingsen}, {Herrnstein}, {Jauncey}, {McCulloch}, {Moran}, {Norris}, {Reynolds}, \& {Tzioumis}}]{greenhill_2003}
{Greenhill}, L.~J., {Booth}, R.~S., {Ellingsen}, S.~P., {et~al.} 2003, \apj, 590, 162, \dodoi{10.1086/374862}

\bibitem[{{Hawley} {et~al.}(1995){Hawley}, {Gammie}, \& {Balbus}}]{hawley_1995}
{Hawley}, J.~F., {Gammie}, C.~F., \& {Balbus}, S.~A. 1995, \apj, 440, 742, \dodoi{10.1086/175311}

\bibitem[{{Hawley} \& {Krolik}(2018)}]{hawley_krolik_2018}
{Hawley}, J.~F., \& {Krolik}, J.~H. 2018, \apj, 866, 5, \dodoi{10.3847/1538-4357/aadf90}

\bibitem[{{Hawley} \& {Krolik}(2019)}]{hawley_krolik_2019}
---. 2019, \apj, 878, 149, \dodoi{10.3847/1538-4357/ab1f6e}

\bibitem[{{Held} \& {Ogilvie}(2024)}]{held_ogilvie_2024}
{Held}, L.~E., \& {Ogilvie}, G.~I. 2024, \mnras, 535, 3108, \dodoi{10.1093/mnras/stae2487}

\bibitem[{{Ingram} {et~al.}(2016){Ingram}, {van der Klis}, {Middleton}, {Done}, {Altamirano}, {Heil}, {Uttley}, \& {Axelsson}}]{ingram_2016}
{Ingram}, A., {van der Klis}, M., {Middleton}, M., {et~al.} 2016, \mnras, 461, 1967, \dodoi{10.1093/mnras/stw1245}

\bibitem[{{Kaaz} {et~al.}(2023){Kaaz}, {Liska}, {Jacquemin-Ide}, {Andalman}, {Musoke}, {Tchekhovskoy}, \& {Porth}}]{kaaz_2023}
{Kaaz}, N., {Liska}, M. T.~P., {Jacquemin-Ide}, J., {et~al.} 2023, \apj, 955, 72, \dodoi{10.3847/1538-4357/ace051}

\bibitem[{{King} {et~al.}(2007){King}, {Pringle}, \& {Livio}}]{king_2007}
{King}, A.~R., {Pringle}, J.~E., \& {Livio}, M. 2007, \mnras, 376, 1740, \dodoi{10.1111/j.1365-2966.2007.11556.x}

\bibitem[{{Kotze} \& {Charles}(2012)}]{kotze_2012}
{Kotze}, M.~M., \& {Charles}, P.~A. 2012, \mnras, 420, 1575, \dodoi{10.1111/j.1365-2966.2011.20146.x}

\bibitem[{{Krolik} \& {Hawley}(2015)}]{krolik_hawley_2015}
{Krolik}, J.~H., \& {Hawley}, J.~F. 2015, \apj, 806, 141, \dodoi{10.1088/0004-637X/806/1/141}

\bibitem[{{Larwood} {et~al.}(1996){Larwood}, {Nelson}, {Papaloizou}, \& {Terquem}}]{larwood_1996}
{Larwood}, J.~D., {Nelson}, R.~P., {Papaloizou}, J.~C.~B., \& {Terquem}, C. 1996, \mnras, 282, 597, \dodoi{10.1093/mnras/282.2.597}

\bibitem[{{Lawrence}(2018)}]{lawrence_2018}
{Lawrence}, A. 2018, Nature Astronomy, 2, 102, \dodoi{10.1038/s41550-017-0372-1}

\bibitem[{{Liska} {et~al.}(2018){Liska}, {Hesp}, {Tchekhovskoy}, {Ingram}, {van der Klis}, \& {Markoff}}]{liska_2018}
{Liska}, M., {Hesp}, C., {Tchekhovskoy}, A., {et~al.} 2018, \mnras, 474, L81, \dodoi{10.1093/mnrasl/slx174}

\bibitem[{{Liska} {et~al.}(2021){Liska}, {Hesp}, {Tchekhovskoy}, {Ingram}, {van der Klis}, {Markoff}, \& {Van Moer}}]{liska_2021}
---. 2021, \mnras, 507, 983, \dodoi{10.1093/mnras/staa099}

\bibitem[{{Liska} {et~al.}(2019){Liska}, {Tchekhovskoy}, {Ingram}, \& {van der Klis}}]{liska_2019}
{Liska}, M., {Tchekhovskoy}, A., {Ingram}, A., \& {van der Klis}, M. 2019, \mnras, 487, 550, \dodoi{10.1093/mnras/stz834}

\bibitem[{{Liska} {et~al.}(2023){Liska}, {Kaaz}, {Musoke}, {Tchekhovskoy}, \& {Porth}}]{liska_kaaz_2023}
{Liska}, M.~T.~P., {Kaaz}, N., {Musoke}, G., {Tchekhovskoy}, A., \& {Porth}, O. 2023, \apjl, 944, L48, \dodoi{10.3847/2041-8213/acb6f4}

\bibitem[{{Liska} {et~al.}(2022){Liska}, {Chatterjee}, {Issa}, {Yoon}, {Kaaz}, {Tchekhovskoy}, {van Eijnatten}, {Musoke}, {Hesp}, {Rohoza}, {Markoff}, {Ingram}, \& {van der Klis}}]{HAMR}
{Liska}, M.~T.~P., {Chatterjee}, K., {Issa}, D., {et~al.} 2022, \apjs, 263, 26, \dodoi{10.3847/1538-4365/ac9966}

\bibitem[{{Lodato} \& {Price}(2010)}]{lodato_2010}
{Lodato}, G., \& {Price}, D.~J. 2010, \mnras, 405, 1212, \dodoi{10.1111/j.1365-2966.2010.16526.x}

\bibitem[{{Lodato} \& {Pringle}(2006)}]{lodato_2006}
{Lodato}, G., \& {Pringle}, J.~E. 2006, \mnras, 368, 1196, \dodoi{10.1111/j.1365-2966.2006.10194.x}

\bibitem[{{Marino} {et~al.}(2015){Marino}, {Perez}, \& {Casassus}}]{marino_perez_2015}
{Marino}, S., {Perez}, S., \& {Casassus}, S. 2015, \apjl, 798, L44, \dodoi{10.1088/2041-8205/798/2/L44}

\bibitem[{{Miyoshi} {et~al.}(1995){Miyoshi}, {Moran}, {Herrnstein}, {Greenhill}, {Nakai}, {Diamond}, \& {Inoue}}]{miyoshi_1995}
{Miyoshi}, M., {Moran}, J., {Herrnstein}, J., {et~al.} 1995, \nat, 373, 127, \dodoi{10.1038/373127a0}

\bibitem[{{Musoke} {et~al.}(2023){Musoke}, {Liska}, {Porth}, {van der Klis}, \& {Ingram}}]{gibwa_2023}
{Musoke}, G., {Liska}, M., {Porth}, O., {van der Klis}, M., \& {Ingram}, A. 2023, \mnras, 518, 1656, \dodoi{10.1093/mnras/stac2754}

\bibitem[{{Nealon} {et~al.}(2018){Nealon}, {Dipierro}, {Alexander}, {Martin}, \& {Nixon}}]{nealon_2018}
{Nealon}, R., {Dipierro}, G., {Alexander}, R., {Martin}, R.~G., \& {Nixon}, C. 2018, \mnras, 481, 20, \dodoi{10.1093/mnras/sty2267}

\bibitem[{{Nelson} \& {Papaloizou}(1999)}]{nelson_1999}
{Nelson}, R.~P., \& {Papaloizou}, J. C.~B. 1999, \mnras, 309, 929, \dodoi{10.1046/j.1365-8711.1999.02894.x}

\bibitem[{{Nixon} {et~al.}(2013){Nixon}, {King}, \& {Price}}]{nixon_2013}
{Nixon}, C., {King}, A., \& {Price}, D. 2013, \mnras, 434, 1946, \dodoi{10.1093/mnras/stt1136}

\bibitem[{{Nixon} {et~al.}(2012){Nixon}, {King}, {Price}, \& {Frank}}]{nixon_2012}
{Nixon}, C., {King}, A., {Price}, D., \& {Frank}, J. 2012, \apjl, 757, L24, \dodoi{10.1088/2041-8205/757/2/L24}

\bibitem[{{Novikov} \& {Thorne}(1973)}]{novikov_thorne_1973}
{Novikov}, I.~D., \& {Thorne}, K.~S. 1973, in Black Holes (Les Astres Occlus), 343--450

\bibitem[{{Ogilvie}(1999)}]{ogilvie_1999}
{Ogilvie}, G.~I. 1999, \mnras, 304, 557, \dodoi{10.1046/j.1365-8711.1999.02340.x}

\bibitem[{{Ogilvie}(2000)}]{ogilvie_2000}
---. 2000, \mnras, 317, 607, \dodoi{10.1046/j.1365-8711.2000.03654.x}

\bibitem[{{Ogilvie}(2006)}]{ogilvie_2006}
---. 2006, \mnras, 365, 977, \dodoi{10.1111/j.1365-2966.2005.09776.x}

\bibitem[{{Ogilvie}(2022)}]{ogilvie_2022}
---. 2022, \mnras, 513, 1701, \dodoi{10.1093/mnras/stac939}

\bibitem[{{Ogilvie} \& {Dubus}(2001)}]{ogilvie_dubus_2001}
{Ogilvie}, G.~I., \& {Dubus}, G. 2001, \mnras, 320, 485, \dodoi{10.1046/j.1365-8711.2001.04011.x}

\bibitem[{{Ogilvie} \& {Latter}(2013)}]{ogilvie_latter_2013}
{Ogilvie}, G.~I., \& {Latter}, H.~N. 2013, \mnras, 433, 2420, \dodoi{10.1093/mnras/stt917}

\bibitem[{{Papaloizou} \& {Lin}(1995)}]{papaloizou_lin_1995}
{Papaloizou}, J.~C.~B., \& {Lin}, D.~N.~C. 1995, \apj, 438, 841, \dodoi{10.1086/175127}

\bibitem[{{Papaloizou} \& {Pringle}(1983)}]{papaloizou_1983}
{Papaloizou}, J.~C.~B., \& {Pringle}, J.~E. 1983, \mnras, 202, 1181, \dodoi{10.1093/mnras/202.4.1181}

\bibitem[{{Priedhorsky} \& {Holt}(1987)}]{priedhorsky_1987}
{Priedhorsky}, W.~C., \& {Holt}, S.~S. 1987, \ssr, 45, 291, \dodoi{10.1007/BF00171997}

\bibitem[{{Pringle}(1981)}]{pringle_1981}
{Pringle}, J.~E. 1981, \araa, 19, 137, \dodoi{10.1146/annurev.aa.19.090181.001033}

\bibitem[{{Pringle}(1992)}]{pringle_1992}
---. 1992, \mnras, 258, 811, \dodoi{10.1093/mnras/258.4.811}

\bibitem[{{Ressler} {et~al.}(2017){Ressler}, {Tchekhovskoy}, {Quataert}, \& {Gammie}}]{ressler_2017}
{Ressler}, S.~M., {Tchekhovskoy}, A., {Quataert}, E., \& {Gammie}, C.~F. 2017, \mnras, 467, 3604, \dodoi{10.1093/mnras/stx364}

\bibitem[{{Ricci} \& {Trakhtenbrot}(2023)}]{ricci_2023}
{Ricci}, C., \& {Trakhtenbrot}, B. 2023, Nature Astronomy, 7, 1282, \dodoi{10.1038/s41550-023-02108-4}

\bibitem[{{Shakura} \& {Sunyaev}(1973)}]{ss_73}
{Shakura}, N.~I., \& {Sunyaev}, R.~A. 1973, \aap, 24, 337

\bibitem[{{Sironi} \& {Tran}(2024)}]{sironi_2024}
{Sironi}, L., \& {Tran}, A. 2024, arXiv e-prints, arXiv:2402.13317, \dodoi{10.48550/arXiv.2402.13317}

\bibitem[{{Smale} \& {Lochner}(1992)}]{smale_1992}
{Smale}, A.~P., \& {Lochner}, J.~C. 1992, \apj, 395, 582, \dodoi{10.1086/171678}

\bibitem[{{Sorathia} {et~al.}(2013){Sorathia}, {Krolik}, \& {Hawley}}]{sorathia_2013}
{Sorathia}, K.~A., {Krolik}, J.~H., \& {Hawley}, J.~F. 2013, \apj, 777, 21, \dodoi{10.1088/0004-637X/777/1/21}

\bibitem[{{Stella} \& {Vietri}(1998)}]{stella_vietri_1998}
{Stella}, L., \& {Vietri}, M. 1998, \apjl, 492, L59, \dodoi{10.1086/311075}

\bibitem[{{Stolker} {et~al.}(2016){Stolker}, {Dominik}, {Avenhaus}, {Min}, {de Boer}, {Ginski}, {Schmid}, {Juhasz}, {Bazzon}, {Waters}, {Garufi}, {Augereau}, {Benisty}, {Boccaletti}, {Henning}, {Langlois}, {Maire}, {M{\'e}nard}, {Meyer}, {Pinte}, {Quanz}, {Thalmann}, {Beuzit}, {Carbillet}, {Costille}, {Dohlen}, {Feldt}, {Gisler}, {Mouillet}, {Pavlov}, {Perret}, {Petit}, {Pragt}, {Rochat}, {Roelfsema}, {Salasnich}, {Soenke}, \& {Wildi}}]{stolker_2016}
{Stolker}, T., {Dominik}, C., {Avenhaus}, H., {et~al.} 2016, \aap, 595, A113, \dodoi{10.1051/0004-6361/201528039}

\bibitem[{{White} {et~al.}(2020){White}, {Dexter}, {Blaes}, \& {Quataert}}]{white_2020}
{White}, C.~J., {Dexter}, J., {Blaes}, O., \& {Quataert}, E. 2020, \apj, 894, 14, \dodoi{10.3847/1538-4357/ab8463}

\bibitem[{{White} {et~al.}(2019){White}, {Quataert}, \& {Blaes}}]{white_2019}
{White}, C.~J., {Quataert}, E., \& {Blaes}, O. 2019, \apj, 878, 51, \dodoi{10.3847/1538-4357/ab089e}

\bibitem[{{Zaw} {et~al.}(2020){Zaw}, {Rosenthal}, {Katkov}, {Gelfand}, {Chen}, {Greenhill}, {Brisken}, \& {Noori}}]{zaw_2020}
{Zaw}, I., {Rosenthal}, M.~J., {Katkov}, I.~Y., {et~al.} 2020, \apj, 897, 111, \dodoi{10.3847/1538-4357/ab9944}

\bibitem[{{Zhao} {et~al.}(2018){Zhao}, {Braatz}, {Condon}, {Lo}, {Reid}, {Henkel}, {Pesce}, {Greene}, {Gao}, {Kuo}, \& {Impellizzeri}}]{zhao_2018}
{Zhao}, W., {Braatz}, J.~A., {Condon}, J.~J., {et~al.} 2018, \apj, 854, 124, \dodoi{10.3847/1538-4357/aaa95c}

\bibitem[{{Zhu}(2019)}]{zhu_2019}
{Zhu}, Z. 2019, \mnras, 483, 4221, \dodoi{10.1093/mnras/sty3358}

\end{thebibliography}

\end{document}